\documentclass[openacc]{rstransa}
\usepackage{subfigure}
\usepackage{pifont}
\usepackage[dvipsnames]{xcolor}
\usepackage{setspace}
\usepackage{graphicx}
\definecolor{color_ripple_carry}{HTML}{EA4335} 
\definecolor{color_carry_look_ahead}{HTML}{4285F4} 
\definecolor{color_hybrid}{HTML}{34A853} 
\definecolor{color_carry_save}{HTML}{F97D53} 

\definecolor{color_quantum_adder}{HTML}{4285F4} 
\definecolor{color_classical_adder}{HTML}{EA4335} 

\definecolor{color_restoring}{HTML}{3C93C2} 
\definecolor{color_non_restoring}{HTML}{226E9C} 
\definecolor{color_long_division}{HTML}{0D4A70} 

\definecolor{color_granlund_montgomery}{HTML}{204964} 

\definecolor{color_schoolbook}{HTML}{AF58BA} 
\definecolor{color_shift_and_add}{HTML}{FF571F} 
\definecolor{color_karatsuba}{HTML}{EA4335} 
\definecolor{color_toom_cook}{HTML}{4285F4} 
\definecolor{color_windowed}{HTML}{FFC61E} 
\definecolor{color_schonhage_strassen}{HTML}{7065BD} 
\definecolor{color_wallace_tree}{HTML}{34A853} 

\definecolor{color_toom_cook_2}{HTML}{46AEA0} 
\definecolor{color_toom_cook_2_5}{HTML}{089099} 
\definecolor{color_toom_cook_3}{HTML}{0071BB} 
\definecolor{color_toom_cook_4}{HTML}{045275} 
\definecolor{color_toom_cook_8}{HTML}{003147} 

\definecolor{color_problem_aware_designs}{HTML}{34A853} 
\definecolor{color_convolution_theorem}{HTML}{EA4335} 

\definecolor{color_goldschmidt}{HTML}{EA4335} 
\definecolor{color_newton_raphson}{HTML}{F97B4F} 

\definecolor{color_clifford_plus_t}{HTML}{4285F4} 
\definecolor{color_qft}{HTML}{EA4335} 
\definecolor{color_both}{HTML}{34A853} 

\usepackage{braket}

\titlehead{Research}

\begin{document}
\title{A Comprehensive Study of Quantum Arithmetic Circuits}

\author{
Siyi Wang$^{1}$, Xiufan Li$^{2}$, Wei Jie Bryan Lee$^{1}$, Suman Deb$^{1}$, Eugene Lim$^{1}$ and Anupam Chattopadhyay$^{1}$}

\address{$^{1}$College of Computing and Data Science, Nanyang Technological University,
50 Nanyang Avenue, Singapore 639798.\\$^{2}$Centre for Quantum Technologies, National University of Singapore, 3 Science Drive 2, Singapore 117543.}

\subject{Quantum Computing}

\keywords{Quantum Computing, 
Quantum Arithmetic, 
Efficiency Optimization,
Quantum Simulation, 
Quantum Hardware.}

\corres{Siyi Wang\\
\email{SIYI002@e.ntu.edu.sg}}

\begin{abstract}

In recent decades, the field of quantum computing has experienced remarkable progress. This progress is marked by the superior performance of many quantum algorithms compared to their classical counterparts, with Shor's algorithm serving as a prominent illustration. Quantum arithmetic circuits, which are the fundamental building blocks in numerous quantum algorithms, have attracted much attention. Despite extensive exploration of various designs in the existing literature, researchers remain keen on developing novel designs and improving existing ones.

In this review article, we aim to provide a systematically organized and easily comprehensible overview of the current state-of-the-art in quantum arithmetic circuits. Specifically, this study covers fundamental operations such as addition, subtraction, multiplication, division and modular exponentiation. We delve into the detailed quantum implementations of these prominent designs and evaluate their efficiency considering various objectives. We also discuss potential applications of presented arithmetic circuits and suggest future research directions.
\end{abstract}


\begin{fmtext}

\end{fmtext}
\maketitle


\section{Introduction\label{sec: Intro}}
\subsection{Background}
Quantum arithmetic is a fundamental building block of quantum computing, involving the design of efficient arithmetic modules using basic quantum operations. These modules include, but are not limited to addition, subtraction, multiplication, division, and modular exponentiation. These quantum arithmetic modules play a vital role in many important quantum algorithms, such as Shor's Algorithm~\cite{Shor}.

\begin{figure}[h]
    \centering
    \includegraphics[width=0.86\textwidth]{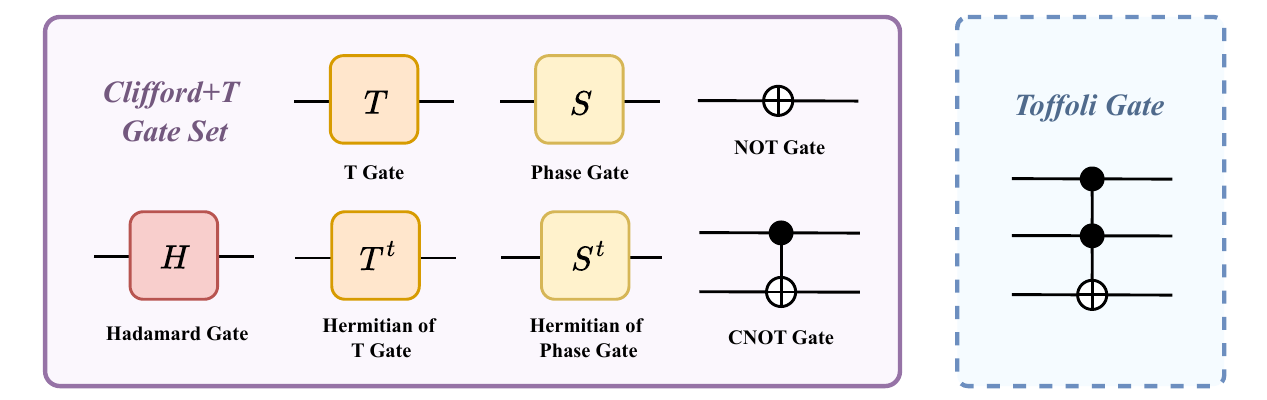}
    \caption{Clifford+T gate set.}
    \label{figClifford+T}
\end{figure}

In the field of quantum computing, one main design method utilizes the Clifford+T gate set. Circuits designed with this method can leverage advanced fault-tolerant architectures based on quantum error-correcting codes (QECC), such as the surface code, thereby alleviating major challenges like the fragility of quantum states, short coherence times, and vulnerability to external noise. In this paper, to standardize and compare various quantum designs, we discuss the T gate cost primarily at the level of the Toffoli gate.
This method implements quantum circuits using Clifford gates (including CNOT, Hadamard, and Phase gates) combined with T gates, which are detailed in Figure~\ref{figClifford+T}. Although Clifford gates are relatively simple to implement and have lower error rates, T gates are complex and costly, making the optimization of T gate usage crucial for designing efficient quantum circuits.

\vspace{-8pt}
\vspace{-20pt}

\begin{figure}[h]
    \centering
    \includegraphics[width=0.86\textwidth]{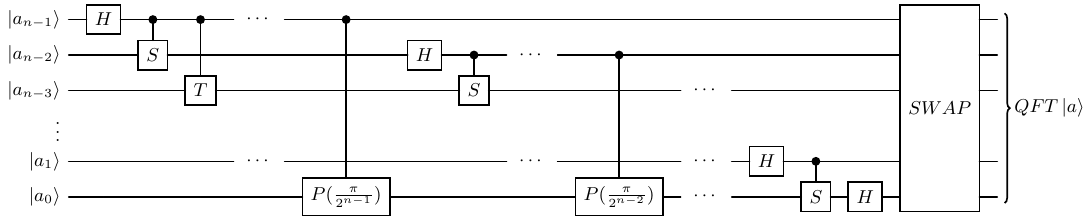}
    \caption{Quantum Fourier Transformation to a $n$ qubit quantum state $\ket{a}:=\ket{a_{n-1}a_{n-2}\cdots a_1a_0}$.}
    \label{figQFT}
\end{figure}

Another mainstream arithmetic framework is based on quantum Fourier transformation (QFT). Relying on the principles of classical Fourier transform, QFT corresponds to the quantum counterpart that maps binary basis to Fourier basis~\cite{nielsen2012quantum}. QFT circuit has a fixed structure as is shown in Figure~\ref{figQFT}, where $P(\phi):=\ket{0}\bra{0} + e^{i\phi}\ket{1}\bra{1}$ denotes the phase gate that rotate the quantum state around Z-axis in the Block sphere with an angle of $\phi \in [0, 2\pi)$.

\vspace{-10pt}

\begin{figure}[h]
    \centering
    \includegraphics[width=0.52\textwidth]{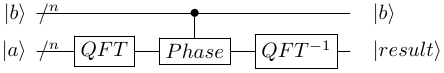}
    \caption{Architecture of QFT-based arithmetic circuits.}
    \label{figQFTArithmeticArchitecture}
\end{figure}

Comprehensive reviews of QFT-based arithmetic frameworks can be found in~\cite{ruizperez2017quantum, atchadeadelomou2023efficient}. QFT-based quantum arithmetic circuit typically begins with a QFT block converting the input state to the frequency domain, performing arithmetic operations using controlled phase gates, and then recovering the state to the original domain through inverse Fourier transform. This approach leverages efficient operations in the Fourier basis to simplify arithmetic operations, particularly in scenarios requiring multiplication and exponentiation. Figure~\ref{figQFTArithmeticArchitecture} shows an illustrative diagram of the schematic of QFT-based quantum arithmetic circuit.

\vspace{-20pt}

Quantum arithmetic circuits hold immense potential for accelerating large-scale quantum algorithms. For instance, in Shor's algorithm, efficient quantum arithmetic circuits can significantly reduce computational complexity and speed up the factorization process. Similarly, the efficiency of quantum arithmetic circuits is directly related to the overall performance of the algorithm.
Quantum arithmetic circuits have garnered widespread attention, not only in academia but also in the industry. Leading technology companies such as Google and IBM have invested significant resources and research interest in this field. These companies aim to advance the practical application and performance of quantum computers through the study and application of quantum arithmetic circuits.
As quantum computing technology advances, the demand for quantum arithmetic circuits will continue to grow. Their broad application prospects in accelerating quantum algorithms, quantum communication, and quantum cryptography herald the arrival of the quantum computing era. Research on quantum arithmetic circuits is not only theoretically significant but also poised to bring substantial technological breakthroughs and innovative applications.

\subsection{Motivation and Contribution}
With the continuous advancement of quantum computers, the importance of quantum arithmetic circuits is increasing. Consequently, substantial research efforts are focused on studying more efficient arithmetic circuits and improving quantum-based techniques to develop these critical units, meeting application requirements in the quantum field.

In this domain, several surveys have been conducted. For instance, Takahashi~\cite{takahashi2009quantum} summarizes efficient quantum circuits for basic arithmetic operations used in Shor's factoring algorithm, focusing on quantum addition, comparison, and the quantum Fourier transform for addition. Additionally, F. Orts et al.\cite{Add_Overview} have conducted a detailed investigation of one of the most essential quantum arithmetic circuits—the quantum reversible adder. Moreover, Sousa\cite{Overview_Nonconventional} extensively reviewed various non-conventional computer arithmetic approaches, including quantum arithmetic circuits.

This paper aims to provide a detailed survey of the latest developments in basic quantum arithmetic circuits by examining two mainstream designs: those based on the Clifford+T set and those based on the QFT. Compared to previous studies, our paper offers a more comprehensive coverage of recent advancements in various basic arithmetic operations, providing detailed technical background and implementation details.

Through this work, we aim to provide researchers and engineers with a thorough reference to better understand and apply the latest quantum arithmetic designs, thereby advancing the field of quantum computing.

\subsection{Organization}
\begin{figure}[ht!]
    \centering
    \includegraphics[width=0.52\textwidth]{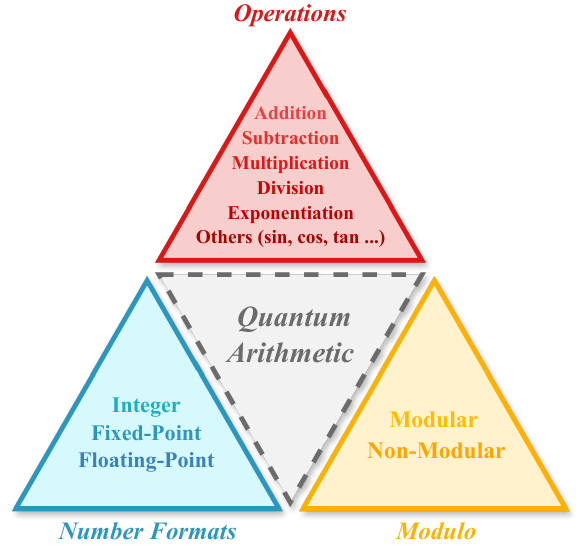}
    \caption{Three main aspects of quantum arithmetic designs discussed in this paper.}
    \label{quantum_arithmetic}
\end{figure}

In this paper, we provide a comprehensive review of the latest designs in quantum arithmetic circuits, delving into various performance metrics and data formats. Quantum arithmetic circuits primarily rely on two mainstream foundational methods: one based on Clifford+T gates and the other on QFT. Our work thoroughly explores both of these primary design approaches in various aspects, as detailed in Figure~\ref{quantum_arithmetic}.

First, in Section~\ref{sec:Number}, we discuss data formats, categorizing all arithmetic operations into three data types: integer, fixed-point, and floating-point. 
Next, in Section~\ref{sec: Eval}, we explore the primary performance evaluation metrics involved in quantum computing. We focus on the metrics that can assess the performance of Clifford+T-based quantum circuits and QFT-based quantum circuits, respectively.
Then, in Sections~\ref{sec: add},~\ref{sec: sub},~\ref{sec: mul},~\ref{sec: div}, and~\ref{sec: mod}, we discuss existing mainstream designs for quantum addition, subtraction, multiplication, division, and modular exponentiation. Each section covers both Clifford+T and QFT designs, providing detailed analyses of their implementation and performance.
In Section~\ref{sec: app}, we elucidate the applications of quantum arithmetic circuits in large-scale quantum algorithms, highlighting their significance and potential impact in practical scenarios.
Finally, in Section~\ref{sec: con}, we summarize the paper, reviewing the key findings and suggesting directions for future research.

\section{Number Formats \label{sec:Number}}
When designing quantum arithmetic circuits, it is essential to consider number formatting. In this work, we focus on three main number formats for arithmetic designs as follows.
\subsection{Integer}
Integers are whole numbers without any fractional components and can be represented in various sizes, such as the commonly used 8-bit, 32-bit, or 64-bit formats. In both classical and quantum computing, integer operations are simple and efficient, making them widely applicable for fundamental tasks such as indexing, counting, and modular arithmetic. As a result, most arithmetic circuit designs in quantum computing focus on integer operations.

\subsection{Fixed-Point}
Fixed-point numbers have a whole part and a fractional part, separated by a decimal point. This format allows for storing values with a fixed range and precision. Fixed-point arithmetic involves operations similar to those in integer arithmetic, but with the consideration of the position of the decimal point.

Compared to integer formats, fixed-point numbers have received less attention in the quantum computing field. Typically, the designs for quantum fixed-point arithmetic are very similar to those for integer arithmetic circuits with the main difference lying in the representation block for fixed-point numbers. For example, several fixed-point quantum arithmetic designs proposed by Zhang~\cite{fixed_point_Zhang} demonstrate structures very similar to corresponding integer operations.

\subsection{Floating-Point}
\begin{figure*}[!ht]
    \centering
    \includegraphics [width=0.86\textwidth]{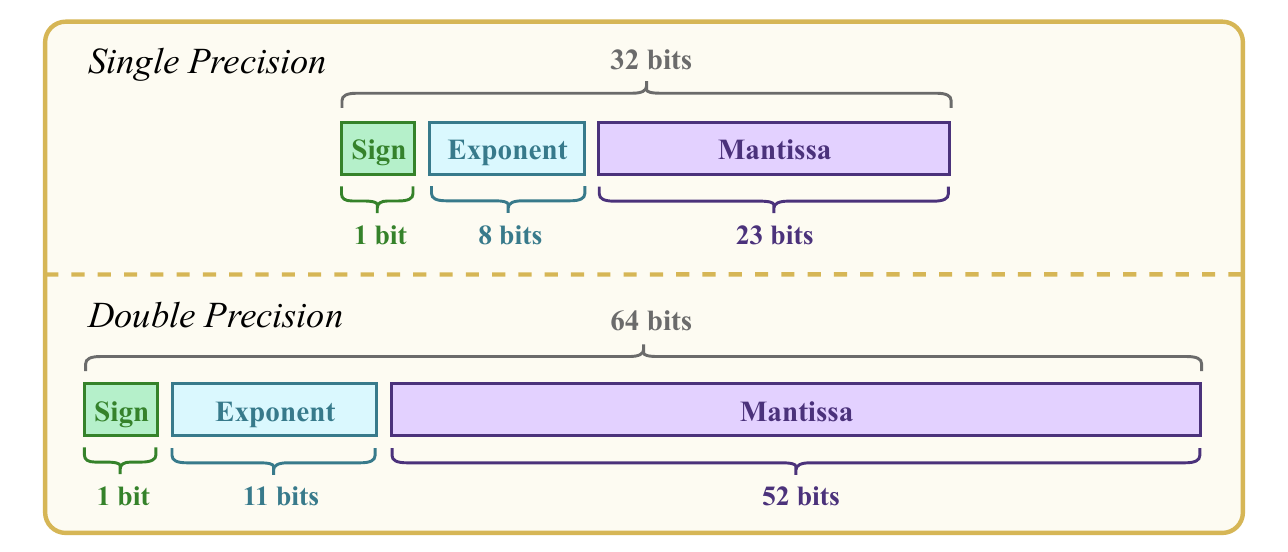}
    \caption{IEEE-754 Floating-Point standard.}\label{IEEE-754 Standard}
\end{figure*}

Floating-point numbers each contain sign, exponent, and mantissa components. The representation of floating-point numbers offers the advantage of dynamic precision, resulting in a much larger range of values. Unlike fixed-point numbers, the precision of floating-point values can be dynamic and can be changed by manipulating the exponent and mantissa bits, while still being stored in fixed lengths. This property is crucial as it allows industry experts to design standardized number formats for computing, such as IEEE-754 (Figure~\ref{IEEE-754 Standard}).

\vspace{-15pt}

In the realm of quantum computing, floating-point arithmetic circuits are often extensions of integer designs, primarily involving specialized modules for floating-point representation and processing. Here, we present a comprehensive review of prominent research on quantum floating-point arithmetic designs.

In 2013, Nguyen et al.~\cite{FP_Nguyen} introduced a space-efficient design for a reversible floating-point adder based on the IEEE-754 single-precision floating-point format, which is particularly well-suited for binary quantum computation. The proposed adder addresses the lack of existing designs for reversible floating-point adders at that time, providing a crucial advancement in the field. Following this, Jain et al.~\cite{FP_Jain} proposed designs for reversible floating-point Ripple-Carry addition, multiplication, and division, all based on integer arithmetic blocks. Their work also utilizes the IEEE-754 standard for floating-point representation. 

In 2018, H{\"a}ner et al.~\cite{floating_Thomas} introduced quantum circuits for floating-point addition and multiplication using two distinct methodologies: automated circuit generation from classical Verilog implementations and manual circuit generation with optimization. In 2021, Gayathri et al.~\cite{FP_Gayathri_2} proposed T-count and T-depth optimized quantum circuits for floating-point addition and multiplication using the IEEE-754 format. The following year, they further expanded their research by presenting a quantum circuit for floating-point square root operations based on the classical Babylonian algorithm~\cite{FP_Gayathri_1}. 

In 2022, Zhao et al.~\cite{FP_Zhao} proposed an enhanced quantum floating-point adder based on the IEEE 754 standard. They categorized quantum floating-point adders into four cases: normal, subnormal, mixed, and special cases, and subsequently designed the quantum modules based on each specific case. One of the main challenges for floating-point arithmetic is the efficient encoding and decoding between classical floating-point data and the corresponding quantum states. 

In 2022, Seidel et al.~\cite{Floating_point_Design} presented an efficient design of unsigned integer arithmetic based on the fact that these evaluations can be written as semi-boolean polynomials. Intriguingly, the authors extended the framework to floating point arithmetic, where non-integers are encoded such that the mantissa is quantum and the exponent is a classically known number, with the bit shift operations that map the non-integer coefficients to the corresponding unsigned integers to suit the semi-boolean polynomials. For the circuit design, QFT and its inverse are utilized as subroutines to construct the bit shift operators and reverse the data to the exact floating points before measuring the outcomes. Combining the bit shift procedure and the in-place operations, this design method provides a speed-up by the parallelization of the gate executions over semi-boolean polynomials.

In brief, both quantum fixed-point and floating-point arithmetic designs heavily rely on their respective quantum integer designs. Therefore, for conciseness, we will primarily focus on integer number formats in the following sections.

\section{Evaluation Metrics\label{sec: Eval}}
While there are several other evaluation metrics, such as Quantum Volume~\cite{Quantum_Volume}, Circuit Layer
Operations Per Second (CLOPS)~\cite{CLOPS}, Layer Fidelity~\cite{Layer_Fidelity}, and Error per Layered Gate (EPLG)~\cite{EPLG}, we focus solely on the most relevant ones in this work. Specifically, we emphasize two main categories of designs. For each category, we utilize corresponding metrics to evaluate the efficiency of the arithmetic designs, as outlined below.

\subsection{Evaluation Metrics for Clifford+T-based Arithmetic Design}

For Clifford+T-based design, we focus on three key metrics. Figure~\ref{Basic_Metrics_Clifford} illustrates a basic example of calculating the primary quantum circuit metrics used in this report. These metrics are outlined as follows:

\begin{figure*}[!ht]
    \centering
    \includegraphics[width=0.6\textwidth]{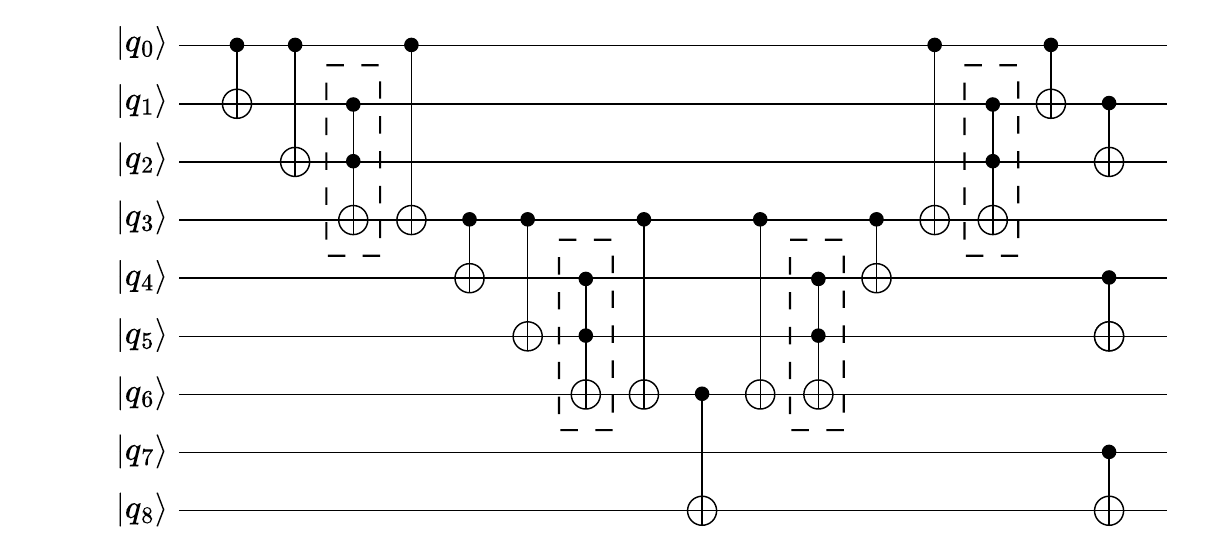}
    \caption{Basic quantum evaluation metrics for Clifford+T-based designs.\label{Basic_Metrics_Clifford} The dotted boxes highlight the Toffoli gates in the circuit. For the circuit shown, the Qubit-Count is $9$, Toffoli-Count is $4$ and Toffoli-Depth is $4$. }
\end{figure*}
\vspace{-15pt}

\begin{itemize}
    \item \textbf{Toffoli-Depth (TD):} It is the most important quantum cost metric as it represents the number of computational layers that include Toffoli gates, which relates to the overall time complexity. Toffoli-Depth can be converted to T-depth using specific decomposition methods, allowing for standardized evaluation.
    \item \textbf{Toffoli-Count (TC):} This metric measures the total number of Toffoli gates within the circuit, providing an estimate of the gate complexity and quantum resource consumption. It can be converted to T-count using specific decomposition methods for standardized comparisons.
    \item \textbf{Qubit-Count (QC):} This metric denotes the total number of qubits required in the specific quantum circuits, correlating with the size of the quantum circuit.
\end{itemize}

\subsection{Evaluation Metrics for QFT-based Arithmetic Design}
\begin{figure}[!ht]
    \centering
    \includegraphics[width=0.9\textwidth]{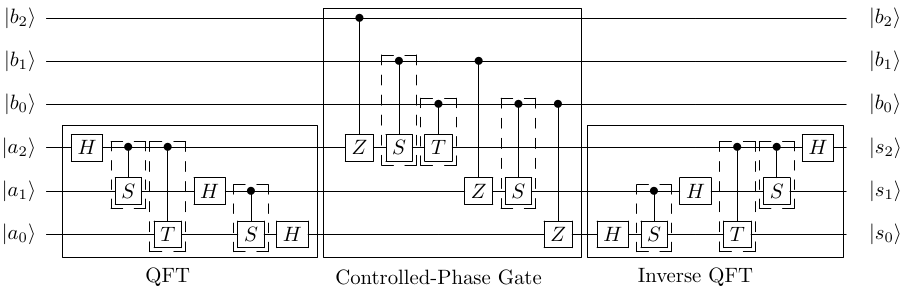}
    \caption{Basic quantum evaluation metrics for QFT-based designs. For $3$-qubit adder, the QFT-Count is $2$ and the Non-Clifford-CP-Count is $3$. Note that the Non-Clifford-CP-Count doesn't include controlled phase gates in the QFT blocks.}
    \label{figQFTMetrics}
\end{figure}

For QFT-based design, the elementary gate set consists of the Hadamard gate and the controlled phase gate ($CP(\phi)$). Controlled phase gates form a family of controlled rotational operations over the z-axis on the Bloch sphere, such that $CP(\phi)=\ket{0}\bra{0} \otimes I + \ket{1}\bra{1} \otimes [\ket{0}\bra{0} + e^{i\phi}\ket{1}\bra{1}]$. The identity gate occurs when $\phi=0$, while the controlled Z gate occurs when $\phi=\pi$. Both are in the Clifford set. For non-Clifford gates, such as the controlled S gate when $\phi=\pi / 2$, the controlled T gate when $\phi=\pi / 4$, and the generally-controlled phase gate when the rotation angle is arbitrarily set, more effort is required for experimental realizations. Thus the performance of QFT-based arithmetic circuits depends predominantly on the number of non-Clifford controlled phase gates. In Figure~\ref{figQFTMetrics}, we illustrate a $3$-qubit QFT-based quantum adder, which has $2$ QFT blocks and $3$ non-Clifford controlled phase gates for addition. The total number of non-Clifford controlled phase gates is $9$. To characterize the overall performance and complexity of QFT-based arithmetic algorithms, we summarize the following three metrics.
\vspace*{-30pt}

\begin{itemize}
    \item \textbf{QFT-Count:} Due to the fixed structure of QFT and its inverse, which depend only on the Qubit-Count, the depth and count of non-Clifford controlled phase gates are always quadratic to the Qubit-Count. Therefore, we consider the number of QFT and its inverse as a performance metric. Typically, one block of QFT for an $n$-qubit circuit requires $n(n-1)/2$ non-Clifford controlled phase gates.
    \item \textbf{Non-Clifford-CP-Count:} After QFT, arithmetic operations such as summations and multiplications are performed on the Fourier basis and usually require sequences of additional controlled phased gates. Therefore, we consider the number of non-Clifford controlled phase gates, excluding those in the QFT subroutines, as another performance metric. A preferred algorithm typically has no additional controlled operations in the middle but is instead designed with single-qubit operations.
    \item \textbf{Qubit-Count:} This metric is the same as in Clifford+T design.
\end{itemize}
We point out that sometimes the evaluation metrics of QFT-based designs are connected to those of Clifford+T designs. For example, in Paler's work~\cite{paler2022quantum}, the QFT-based metric of Non-Clifford-CP-Count is linked to the Clifford+T-based metric of Toffoli-Count by merging adjacent Hadamard gates and controlled rotation gates and formulating them with Clifford and Toffoli operations. Related work on the discussions about resource costs for QFT-based algorithms using Clifford+T-based metrics can be found in reference~\cite{kim2018efficient, nam2020approximate, park2022Tcount}.

\section{Quantum Addition\label{sec: add}}
Addition is a fundamental operation that serves as a cornerstone for the functionality of all algorithms built upon it. In classical computing, addition is typically performed by  adders such as Ripple-Carry, Carry-Lookahead \cite{rosenberger1957simultaneous}, Carry-Save \cite{earle1967latched}, Carry-Select, and Carry-Skip. Similarly, Clifford+T-based and QFT-based quantum adders can be classified into these same categories. Here, we focus exclusively on integer addition designs, as discussed in Section~\ref{sec:Number}.







\subsection{Clifford+T-based Designs}
\begin{figure}[ht!]
    \centering
    \includegraphics[width=0.69\textwidth]{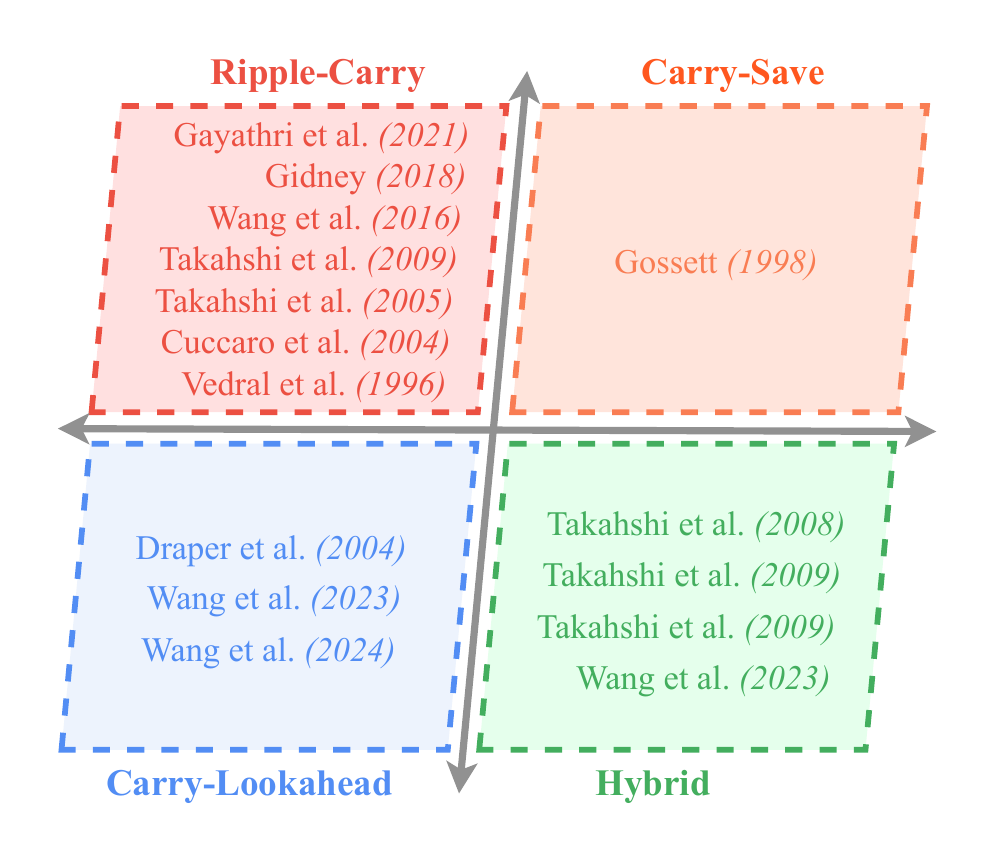}
    \caption{Classification of Clifford+T-based quantum adders based on the four main structures discussed in this paper.}
    \label{addition_clifford_plus_t}
\end{figure}

Table~\ref{table: c+T_Add} presents an overview of prominent Clifford+T-based quantum adders, detailing the publication year, corresponding article, structure, and key innovations for each design. Additionally, Figure~\ref{addition_clifford_plus_t} is provided to illustrate the articles published over the years categorized in the four main structures discussed in this paper for Clifford+T-based quantum adders.

\vspace{-18pt}
\vspace{-25pt}

\begin{table}[!h]
\caption{An overview of prominent Clifford+T-based quantum adders.}
\label{table: c+T_Add}
\centering
\scalebox{0.9}{
\begin{tabular}{llll}
\hline
Year &Article &Structure & Key Innovations \\
\hline
1996 & Vedral et al.~\cite{VBE}&\textcolor{color_ripple_carry}{Ripple-Carry}&
Propose the First Quantum Ripple-Carry Adder 
\\
1998&Gossett~\cite{gossett1998quantum}&\textcolor{orange}{Carry-Save}&Propose the First Quantum Carry-Save Adder\\
2004 & Cuccaro et al.~\cite{Cuccaro}&\textcolor{color_ripple_carry}{Ripple-Carry} & Use Only One Ancilla\\
2004 & Draper et al.~\cite{Draper}&\textcolor{color_carry_look_ahead}{Carry-Lookahead}&Propose the First Quantum Carry-Lookahead Adder\\
2005 & Takahshi et al.~\cite{Takahashi05}& \textcolor{color_ripple_carry}{Ripple-Carry} &Use Zero Ancilla\\
2008 & Takahshi et al.~\cite{Takahashi08}& \textcolor{color_hybrid}{Hybrid}&Design Ripple-Carry \& Carry-Lookahead Adder\\
2009 & Takahshi et al.~\cite{Takahashi09}& \textcolor{color_ripple_carry}{Ripple-Carry}& Reduce the Size \& Depth of Adder with Zero Ancilla \\
2009 & Takahshi et al.~\cite{Takahashi09}& \textcolor{color_hybrid}{Hybrid}& Simplify the Design~\cite{Takahashi08}\\
2009 & Takahshi et al.~\cite{Takahashi09}& \textcolor{color_hybrid}{Modified Hybrid}& Simplify the Design~\cite{Takahashi09} \\
2016 & Wang et al.~\cite{paper4_2016} & \textcolor{color_ripple_carry}{Ripple-Carry} & Simplify the Building Blocks of Ripple-Carry Adder \\
2018 & Gidney~\cite{Gidney2018halvingcostof}&\textcolor{color_ripple_carry}{Ripple-Carry}& Introduce Logical-AND to Reduce the Cost\\
2021 &Gayathri et al.~\cite{paper3_2021}&\textcolor{color_ripple_carry}{Ripple-Carry}&Simplify the Building Blocks of Ripple-Carry Adder\\
2023 &Wang et al.~\cite{wang2023higher} & \textcolor{color_hybrid}{Hybrid}&Simplify the Building Blocks of Hybrid Adder \\
2023 &Wang et al.~\cite{wang2023reducing} & \textcolor{color_carry_look_ahead}{Carry-Lookahead}& Propose Quantum Ling Base Adder\\
2024 &Wang et al.~\cite{wang2024optimal} &\textcolor{color_carry_look_ahead}{Carry-Lookahead} &Explore Parallel Prefix Trees in Quantum \\
\hline
\end{tabular}}
\vspace*{-20pt}
\end{table}

\begin{itemize}
    \item \textbf{Ripple-Carry Structure.}\\
    
The initial design for Clifford+T-based quantum addition dates back to 1996, when Vedral et al.~\cite{VBE} presented an explicit construction of quantum networks capable of performing fundamental arithmetic operations. Among these operations was the initial design of a quantum $n$-bit Ripple-Carry adder, marking a significant milestone in the development of quantum arithmetic.

\begin{doublespace}
    
\end{doublespace}
In 2004, Cucarro et al.~\cite{Cuccaro} introduced a new linear-depth Ripple-Carry quantum addition circuit, following Vedral et al.'s 1996 proposal of a Ripple-Carry adder~\cite{VBE}. While Vedral et al.'s adder~\cite{VBE} required a linear number of ancillary qubits, this newer design~\cite{Cuccaro} relies on just a single ancillary qubit. Additionally, the circuit boasts reduced depth and a smaller gate count compared to earlier Ripple-Carry designs.
In the following year, Takahashi et al.~\cite{Takahashi05} further enhanced the quantum Ripple-Carry adder by developing a quantum circuit for adding two $n$-bit binary numbers without any ancillary qubit.
In 2009, Takahashi et al.~\cite{Takahashi09} further improved the Ripple-Carry adder. This design, with a size of $7n-6$, was smaller than any previous quantum circuit for addition without ancillary qubits.
Additionally, Wang et al.~\cite{paper4_2016} introduced an enhanced linear-depth Ripple-Carry quantum addition circuit. Unlike the previous adders, which typically required at least two Toffoli gates per output bit, this adder utilized only a single Toffoli gate per bit.

\begin{doublespace}
    
\end{doublespace}
In 2018, Gidney~\cite{Gidney2018halvingcostof} introduced a significant design advancement. By devising a novel ''logical-AND'' construction employing four T gates to record the logical-AND of two qubits into an ancilla and zero T gates to subsequently erase the ancilla, This approach effectively halved the cost of current quantum addition.
Moreover, Gayathri et al.\cite{paper3_2021} improved the structure of the quantum full adder and introduced a quantum Ripple-Carry adder based on this improvement. To manage the garbage output during the uncomputing phase of the proposed quantum adder, they used a technique involving circuit reversal using the reversible pebble game in accordance to the Bennet removal scheme\cite{Bennett1973LogicalRO}. As a result, they developed an uncomputing quantum circuit that doesn't rely on Toffoli gates, thereby reducing the excess number of Toffoli gates in the proposed adder circuit.\\

    \item \textbf{Carry-Lookahead Structure.}\\
    
After the introduction of quantum Ripple-Carry adders, Draper et al.~\cite{Draper} introduced an efficient addition circuit inspired by classical Carry-Lookahead arithmetic circuits. This pioneering quantum Carry-Lookahead adder significantly reduces the cost of addition while only marginally increasing the number of required qubits.
In 2023, Wang et al.~\cite{wang2023reducing} introduced the Ling basis in quantum arithmetic. Although this initial quantum Ling adder showed only modest improvement, its significance lies in exploring the potential to alter the propagation and generation base within the quantum Carry-Lookahead design.
In 2024, Wang et al. explored 160 alternative compositions of the carry-propagation structure in their work~\cite{wang2024optimal}, aiming to determine the optimal depth structure for a quantum adder. Through extensive study of these structures, they demonstrated that an exact Toffoli-Depth of $log n + O(1)$ is attainable by implementing the Sklansky prefix tree~\cite{Sklansky} in the quantum Carry-Lookahead adder.\\

    \item \textbf{Carry-Save Structure.}\\
    
The most famous work on Carry-Save structures was introduced by Gossett in 1998~\cite{gossett1998quantum}. Despite its prominence, this work still offers considerable room for improvement. Additionally, Carry-Save structures find extensive application in constructing modular multiplication and modular exponentiation arithmetic circuits, as elaborated in Section~\ref{sec: mod}.\\

    \item \textbf{Hybrid Structure.}\\
    
The Ripple-Carry approach, while reducing the number of qubits, is known to incur large Toffoli-Depth~\cite{VBE, Cuccaro}. Conversely, the Carry-Lookahead approach decreases Toffoli-Depth but requires a larger Qubit-Count~\cite{Draper}. 
Hence, in 2008, the first hybrid structure was introduced to leverage the low Qubit-Count and the low Toffoli-Count advantages of the Ripple-Carry adder as well as the low Toffoli-Depth advantage of the Carry-Lookahead structure. Specifically, Takahashi et al.\cite{Takahashi08} combined a modified version of Draper et al.'s quantum Carry-Lookahead adder\cite{Draper} with parallel applications of Takahashi et al.'s quantum Ripple-Carry adder~\cite{Takahashi05}.
In 2009, Takahashi et al.~\cite{Takahashi09} proposed a generalized and simplified version of the previous hybrid method\cite{Takahashi08} for constructing a logarithmic-depth circuit with a reduced number of qubits. Additionally, they introduced a more efficient version of this hybrid design using unbounded fan-out gates.
In 2023, Wang et al.~\cite{wang2023higher} introduced an efficient quantum Carry-Lookahead adder utilizing a higher radix structure. They achieved this by separating the propagation and summation phases within the Manchester Carry Chain, resulting in a flexible and efficient hybrid addition circuit. This approach allows for various carry paths and sum paths within the quantum higher radix addition framework, enhancing its flexibility and efficiency in different scenarios.
\end{itemize}

\subsection{QFT-based Circuit Designs}
\begin{figure}[ht!]
    \centering
    \includegraphics[width=0.8\textwidth]{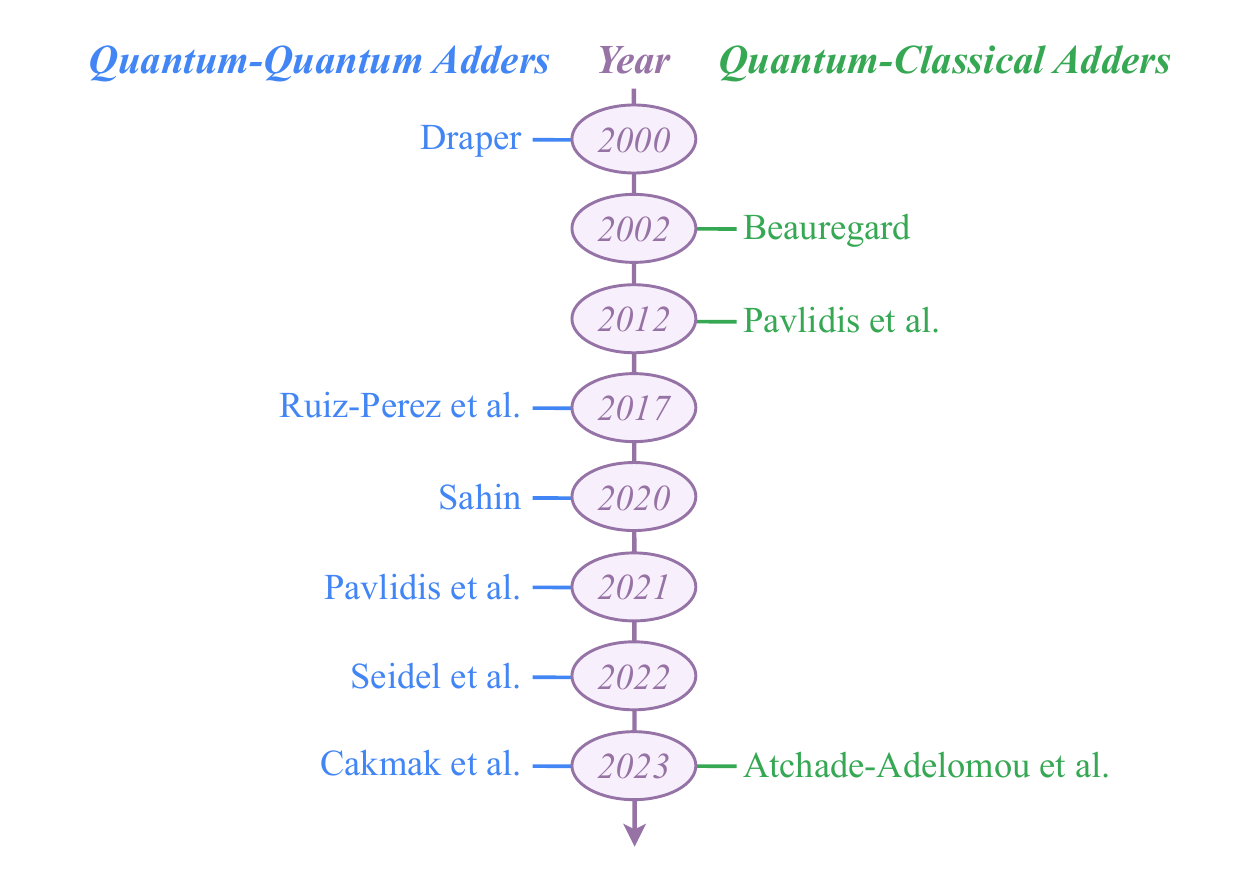}
    \caption{Timeline of QFT-based adders based on structure type.}
    \label{addition_qft_timeline}
\end{figure}

Quantum addition circuits with Fourier transform are categorized into two main structures based on the type of input factors. Advancements in QFT-based adders are listed in Table~\ref{tabQFTAddition}. To better visualize the timeline of Quantum-Quantum Adders and Quantum-Classical Adders published over the years, a timeline is provided in Figure~\ref{addition_qft_timeline}.

\vspace{-10pt}
\vspace{-25pt}

\begin{table}[!h]
    \centering
    \caption{An overview of prominent QFT-based quantum adders.}
    \scalebox{0.8}{
    \begin{tabular}{llll}  \hline
    Year &Article &Structure & Key Innovations \\  \hline
    2000&Draper~\cite{draper2000addition}& \textcolor{color_quantum_adder}{Quantum-Quantum Adder}&Propose the First Quantum-Quantum Adder\\
    2002&Beauregard~\cite{beauregard2002circuit}&\textcolor{color_hybrid}{Quantum-Classical Adder}&Propose the First Quantum-Classical Adder\\
    2012&Pavlidis et al.~\cite{pavlidis2012fast}&\textcolor{color_hybrid}{Quantum-Classical Adder}&Design Multi-Controlled Adder\\
    2017&Ruiz-Perez et al.~\cite{ruizperez2017quantum}&\textcolor{color_quantum_adder}{Quantum-Quantum Adder} &Realize Non-Modular Summation\\
    2020&Sahin~\cite{ahin2020quantum} &\textcolor{color_quantum_adder}{Quantum-Quantum Adder} &Perform Addition on Signed Integers \\
    2021&Pavlidis et al.~\cite{pavlidis2021quantum}&\textcolor{color_quantum_adder}{Quantum-Quantum Adder}&Propose the First QFT-based Qudit Adder\\
    2022&Seidel et al.~\cite{Floating_point_Design}&\textcolor{color_quantum_adder}{Quantum-Quantum Adder}&Propose Adder with Semi-Boolean Polynomials\\
    2023&Atchade-Adelomou et al.~\cite{atchadeadelomou2023efficient}&\textcolor{color_hybrid}{Quantum-Classical Adder}&Summarize Inplace \& Outplace Structure\\
    2023&Cakmak et al.~\cite{cakmak2023QFT}&\textcolor{color_quantum_adder}{Quantum-Quantum Adder}&Execute the Adder in Experimental Demonstration\\
    \hline
    \end{tabular}}
    \label{tabQFTAddition}
\vspace{-15pt}
\end{table}

\begin{itemize}
    \item \textbf{QFT-based Quantum-Quantum Adder.}\\
    
    Given $a$ and $b$ encoded as two $n$-qubit quantum states $\ket{a}$ and $\ket{b}$, the addition circuit of $\ket{a}$ and $\ket{b}$ requires at least $2n$ qubits as inputs. One of the factors, for example $\ket{b}$, is used as the ancilla qubits that control the rotation angles in the phase gates. An illustrative diagram is shown in Figure~\ref{figQFTAdderQuantum}, where the controlled phase gates realize the summation of $b$ on the Fourier basis of $a$. This architecture was first proposed by Draper in 2000~\cite{draper2000addition}. \\
    
    Later, Ruiz-Perez and Garcia-Escartin summarized the design for both modular and non-modular operations by introducing an additional carry qubit and used the adders to compute weighted sums~\cite{ruizperez2017quantum}. In 2020, Sahin applied Ruiz-Perez and Garcia-Escartin's framework for arithmetic operations on signed integers and presented algorithms for calculating two's complement, absolute value, and comparison~\cite{ahin2020quantum}. \\
    
    QFT-based addition was also studied in qudit systems, where data is encoded with a higher dimensional unit~\cite{pavlidis2021quantum}. In 2022, Seidel et al. proposed arithmetic for signed and unsigned integers and floating points using a novel data encoding scheme based on semi-boolean polynomials~\cite{Floating_point_Design}. In their design, integers are encoded as semi-boolean polynomials by a QFT block and decoded by its inverse. Additions with respect to the semi-boolean polynomials are performed using controlled-phase gates in the middle of the circuit. Therefore, we categorize this framework as similar to the QFT-based quantum-quantum adder. \\
    
    In 2023, Cakmak constructed a primitive quantum arithmetic logic unit capable of performing AND and logical NOT-AND operations, and executed it on the IBM quantum platform as an experimental demonstration~\cite{cakmak2023QFT}.
    \vspace*{-6pt}
    \begin{figure}[!h]
        \centering
        \includegraphics[width=\textwidth]{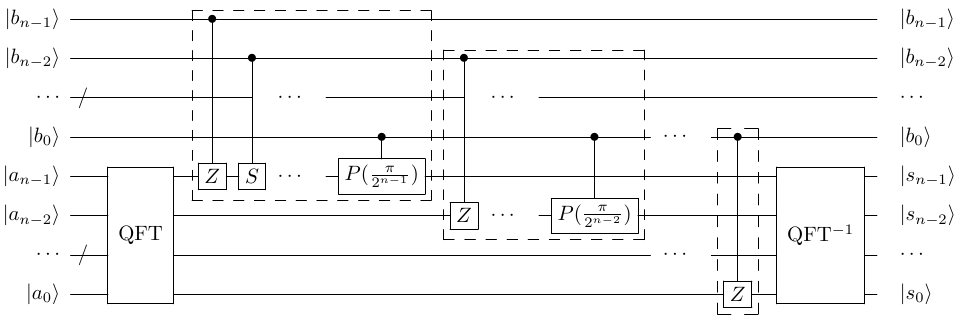}
        \caption{Architecture of QFT-based adder with two quantum states $\ket{a}$ and $\ket{b}$ as inputs.}
        \label{figQFTAdderQuantum}
    \end{figure}
    \vspace{-15pt}
    
    \item \textbf{QFT-based Quantum-Classical Adder.}\\
    
    Another form of the addition circuit takes the inputs of a quantum state and a classical constant. Given the $\ket{a}$ as an initial state and $b$ as a classical value, the QFT-based adder is constructed by precomputing the rotation angles with respect to $b$ and performing single qubit rotations accordingly. The circuit has a reduced number of input qubits. Figure~\ref{figQFTAdderHybrid} illustrates the adder, which encodes the classical summation factor $b$ into rotation angles $\phi(b)$ in the phase gates, realizing the summation of $b$ on the Fourier basis of $a$. This design was first proposed by Beauregard in 2002 as a modified framework of Draper's adder~\cite{beauregard2002circuit}. \\
    
    In 2012, Pavlidis and Gizopoulos proposed the controlled quantum adder and the controlled-controlled quantum adder for arithmetic with a classical constant~\cite{pavlidis2012fast}. As elaborated in the sections below, these two gates are pivotal for high-level arithmetic operations such as multiplication and exponentiation. In 2023, Atchade-Adelomou and Gonzalez further summarized quantum-classical adders with in-place and out-of-place operations, discussing their applications on intermediate-scale platforms~\cite{atchadeadelomou2023efficient}.
    \begin{figure}[!h]
        \centering
        \includegraphics[width=0.56\textwidth]{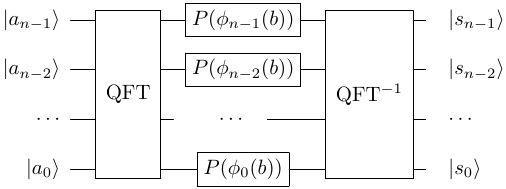}
        \caption{Architecture of QFT-based adder with a quantum state $\ket{a}$ and a classical constant $b$ as inputs.}
        \label{figQFTAdderHybrid}
    \end{figure}
\end{itemize}

\section{Quantum Subtraction \label{sec: sub}}
$N$-bit quantum binary subtractors have structures closely resembling those of quantum adders. We discuss the construction of quantum subtraction circuits for both Clifford+T design and QFT-based designs. These circuits can be categorized into three primary methods, as detailed below.

\subsection{Method 1: Ripple-Borrow Subtraction}
\label{subSecMtd1}
Subtractor circuits, which use two n-bit operands to produce an n-bit result with a borrow-out signal, are less frequently discussed in quantum systems compared to quantum addition circuits. Despite this, quantum subtractor circuits offer significant potential for innovative designs in quantum computing.

\begin{figure*}[!ht]
    \centering
    \includegraphics[width=\textwidth]{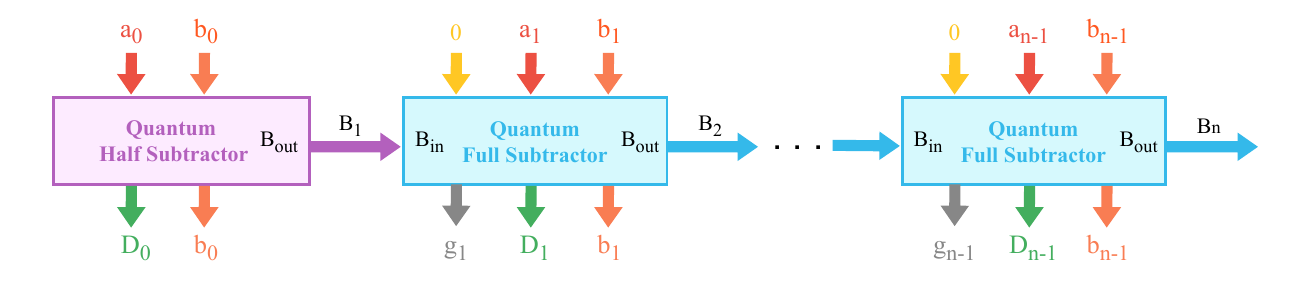}
    \caption{$n$-bit quantum subtractor based on Method 1.\label{Approach1_Sub}}
\end{figure*}

The quantum full subtractor module differs only slightly from the quantum full-adder module, and similarly, the quantum half subtractor module differs only slightly from the quantum half adder module. By using quantum half subtractors and full subtractors, an $n$-bit binary quantum subtractor can be constructed using the Ripple-Borrow approach, as illustrated in Figure~\ref{Approach1_Sub}. Numerous existing quantum works~\cite{cheng2002quantum, thapliyal2009design, thapliyal2016mapping} utilize this method to develop their quantum subtractors. These designs share a similar structure and overall cost with Ripple-Carry Adder circuits, typically requiring a linear Toffoli-Depth.

\vspace{-20pt}

\subsection{Method 2: Subtraction Based on Addition Without Input Carry}
\label{subSecMtd2}
\begin{figure}[ht]
    \centering
    \subfigure[Method 2\label{Approach2_Sub}]{\includegraphics[width=0.5\linewidth]{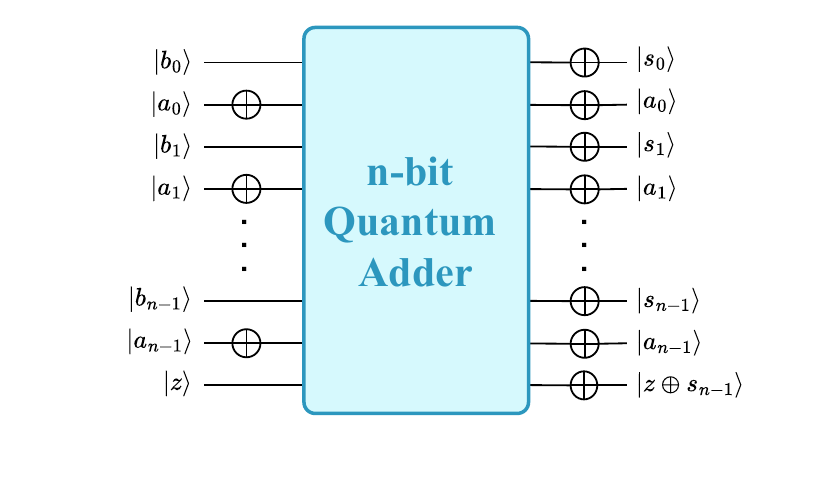}}
    \hfill
    \subfigure[Method 3\label{Approach3_Sub}]{\includegraphics[width=0.48\linewidth]{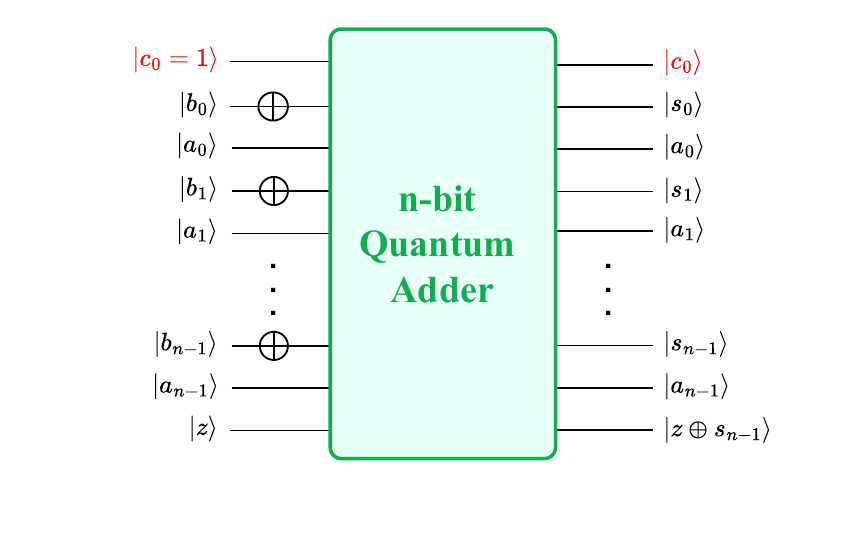}}
    \hfill
    \caption{$n$-bit quantum subtractor.}
\end{figure}

In addition to Method 1, it is possible to construct an n-bit quantum subtractor by making minor modifications to an n-bit quantum adder circuit, thereby using it as a subtractor. These modifications, as exemplified in Formulas \ref{formula: sub0} and \ref{formula: sub1}, can be achieved simply by employing quantum NOT gates.
\begin{align}
&A-B=(A'+ B)'\label{formula: sub0}\\
&A-B=A+ B'+1\label{formula: sub1}
\end{align}
Method 2, illustrated in Formula \ref{formula: sub0}, relies on this principle. Specifically, the design of the quantum $n$-bit subtractor involves complementing the input $a$ at the outset and subsequently complementing both $a$ and the resulting sum at the end. As shown in diagram~\ref{Approach2_Sub}, the proposed subtractor design heavily relies on the structure of the quantum adder used. Notably, these modifications only entail the use of quantum NOT gates, thus eliminating the need for extra Toffoli gates or qubits.
\vspace*{25pt}
\vspace*{-18pt}

For QFT-based designs, Method 2 is the most commonly used principle for constructing a subtractor from an adder. According to the commuting rule of $X$ gate and $P(\phi)$ phase gate, $XP(\phi)X=P(-\phi)$, interpolating quantum NOT gates as a minor modification to the QFT-based adder block is equivalent to a new QFT-based adder block with the same structure but negative rotation angles, as mentioned in Beauregard et al.'s design~\cite{beauregard2002circuit} and Sahin's work~\cite{sahin2020quantum}.

\subsection{Method 3: Subtraction Based on Addition With Input Carry}
\label{subSecMtd3}
In contrast to Method 2, Method 3 leverages an $n$-bit quantum adder with input carry to construct an $n$-bit quantum subtractor. As shown in Formula~\ref{formula: sub1}, this method realizes an $n$-bit quantum subtractor by employing a quantum adder circuit with input carry, as illustrated in diagram \ref{Approach3_Sub}. Specifically, the design process involves complementing the input b at the outset and setting the input carry $c_0$ to 1. In brief, this method is similar to Method 2, with slight differences in the number of NOT gates and Qubit-Count, as shown in Table~\ref{tabCompare}.

\begin{table}[ht!]
\caption{Cost comparison between Method 1 and Method 2 for $n$-bit quantum subtraction.}
\label{tabCompare}
\centering
\begin{tabular}{cccc}
\hline
Subtraction & NOT Gates & Qubit-Count & Carry In \\ \hline
Method 2    & $3n+1$      & $2n+1$   & \textcolor{red}{\ding{55}}     \\
Method 3    & $n$        & $2n+2$   & \textcolor{color_hybrid}{\checkmark}    \\\hline
\end{tabular}
\end{table}

\section{Quantum Multiplication \label{sec: mul}}
In classical computation, advanced multiplication methods such as Shift-and-Add, Karatsuba~\cite{karatsuba1962multiplication}, Wallace Tree~\cite{wallace1964suggestion}, and Toom-Cook~\cite{cook1969minimum} algorithms minimize the number of multiplicative operations, thereby boosting efficiency. 
Fast Fourier Transform (FFT)-based methods, such as the Sch{\"o}nhage-Strassen algorithm~\cite{schonhage1971fast}, further improve the efficiency of large integer multiplication by transforming it into a convolution. 
Similarly, in quantum multiplication, these classical methods are leveraged to perform operations even more efficiently. In this section, we focus solely on binary designs, although there are also several works that concentrate on ternary reversible multipliers~\cite{panahi2021novel,asadi2021towards,faghih2023efficient}.

\subsection{Clifford+T-based Designs}
\begin{figure}[ht!]
    \centering
    \includegraphics[width=\textwidth]{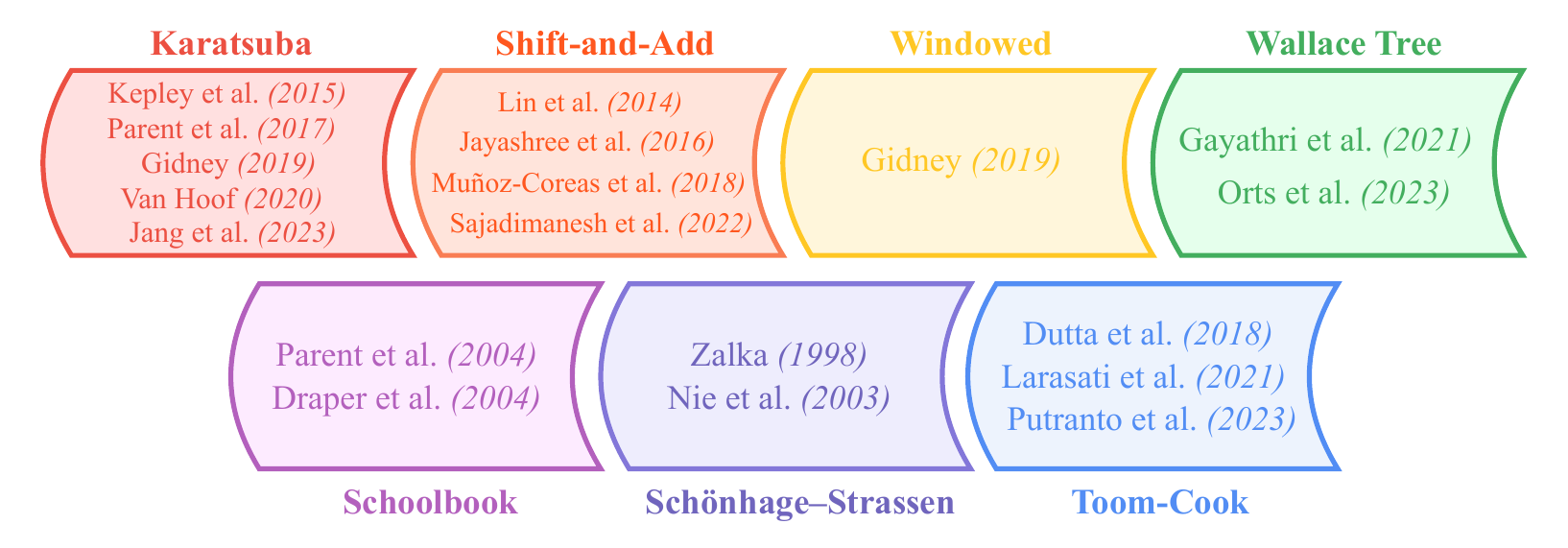}
    \caption{General classification of Clifford+T-based quantum multipliers articles as per structure and sorted by publication year.}
    \label{multiplication_clifford_plus_t}
\end{figure}

Quantum multiplications using Clifford+T-based multipliers are categorized in Table~\ref{tabCliffordPlusTMultiplier}, with details on their respective publication year, corresponding article, structure, and key innovations. A general classification diagram is provided in Figure~\ref{multiplication_clifford_plus_t}, grouping each article based on its structure and sorted by publication year for quantum Clifford+T-based multipliers.

\vspace*{-15pt}
\begin{table}[!h]
\caption{An overview of prominent Clifford+T-based quantum multipliers.}
\label{tabCliffordPlusTMultiplier}
\centering
\scalebox{0.84}{
\begin{tabular}{llll}
\hline
Year &Article &Structure & Key Innovations \\
\hline
1998&Zalka~\cite{zalka1998fast}&\textcolor{color_schonhage_strassen}{Sch{\"o}nhage-Strassen}&Propose the First Sch{\"o}nhage-Strassen Multiplier\\
2004&Parent et al.~\cite{parent2017improved}&\textcolor{color_schoolbook}{Schoolbook}&Use $n$ Plain Adders in Schoolbook Method\\
2004&Draper et al.~\cite{Draper}&\textcolor{color_schoolbook}{Schoolbook}&Use $n$ Carry-Lookahead Adders\\
2014& Lin et al.~\cite{lin2014qlib}&\textcolor{color_shift_and_add}{Shift-and-Add}&Propose the First Shift-and-Add Multiplier\\
2015 &Kepley et al.~\cite{kepley2015quantum}&\textcolor{color_karatsuba}{Karatsuba}&Propose the First Quantum Karatsuba Multiplier\\
2016 & Jayashree et al.~\cite{jayashree2016ancilla}&\textcolor{color_shift_and_add}{Shift-and-Add}& Optimize Ancillas and Eliminate Garbage\\
2017&Parent et al.~\cite{parent2017improved}&\textcolor{color_karatsuba}{Karatsuba}&Optimize Qubit-Count for Karatsuba Multiplier\\
 2018&Dutta et al.~\cite{Toom-Cook-multi}&\textcolor{color_toom_cook}{Toom-Cook 2.5}&Propose the First Quantum Toom-Cook Multiplier\\
 2018&Mu{\~n}oz-Coreas et al.~\cite{munoz2018quantum}&\textcolor{color_shift_and_add}{Shift-and-Add}& Optimize Toffoli-Count and Toffoli-Depth\\
 2019&Gidney~\cite{gidney2019asymptotically}&\textcolor{color_karatsuba}{Karatsuba}& Use Inline-Mutation Calls to Optimize Qubit-Count\\
 2019& Gidney~\cite{gidney2019windowed}&\textcolor{color_windowed}{Windowed}&Introduce Lookup Tables\\
 2020&Van Hoof~\cite{van2019space}&\textcolor{color_karatsuba}{Karatsuba}& Optimize Qubit-Count for Karatsuba Multiplier\\
 2021&Larasati et al.~\cite{larasati2021quantum}&\textcolor{color_toom_cook}{Toom-Cook 3}&Optimize Qubit-Count and Toffoli-Depth\\
 2021& Gayathri et al.~\cite{gayathri2021t}&\textcolor{color_wallace_tree}{Wallace Tree}&Propose the First Quantum Wallace Tree Multiplier\\
 2022 & Sajadimanesh et al.~\cite{sajadimanesh2022practical}&\textcolor{color_shift_and_add}{Shift-and-Add}&Adapt the Design for Quantum Devices \\
 2023&  Putranto et al.~\cite{putranto2023space}& \textcolor{color_toom_cook}{Toom-Cook 8}&Optimize Space \& Time Cost for Toom-Cook Design\\
 2023& Orts et al.~\cite{orts2023improving}&\textcolor{color_wallace_tree}{Wallace Tree}&Support Arbitrary Size Multiplication\\
 2023&Nie et al.~\cite{nie2023quantum}&\textcolor{color_schonhage_strassen}{Sch{\"o}nhage-Strassen}&Improve Sch{\"o}nhage-Strassen Multiplier\\
 2023& Jang et al.~\cite{jang2023quantum}&\textcolor{color_karatsuba}{Karatsuba}&Reduce the Toffoli-Depth to One\\
\hline
\end{tabular}}
\vspace*{-10pt}
\end{table}

\begin{itemize}
    \item \textbf{Schoolbook.}\\
    
    A simple approach to integer multiplication is to reduce it to addition using $n$ adders. This method, known as the schoolbook method, was applied in both Draper et al.\cite{Draper} and Parent et al.\cite{parent2017improved}'s papers. While this early methodology features a straightforward structure, it is not efficient in terms of Qubit-Count, Toffoli-Count, or Toffoli-Depth.\\

    \item \textbf{Shift-and-Add.}\\
    
    Shift-and-Add multiplication is similar to the traditional  grade-school multiplication algorithm. 
    In the quantum module library QLib, proposed by Lin et al.~\cite{lin2014qlib}, quantum modules of various sizes and specifications are available for benchmarking quantum logic and physical synthesis. Although the overall cost is high, the multiplier within QLib represents an initial quantum design based on the Shift-and-Add multiplication method.\\
    
    In 2016, Jayashree et al.\cite{jayashree2016ancilla} proposed a quantum multiplier based on a modified version of the Shift-and-Add method. This design uses reversible logic to create a garbage-free and ancilla qubit-optimized quantum integer multiplier. It addresses the issue of redundant garbage qubits in existing reversible quantum integer multipliers. Specifically, it reduces the required ancilla qubits from $3N+1$, as in the design by Lin et al.\cite{lin2014qlib}, to $2N+1$ for $N$-bit quantum multiplication.\\
    
    In 2018, Mu{\~n}oz-Coreas et al.\cite{munoz2018quantum} introduced a T-count optimized quantum integer multiplier using the Shift-and-Add method. Based on a novel quantum conditional adder circuit, this design achieved a T-count optimized quantum circuit, requiring only $4N + 1$ qubits and generating no garbage outputs. This design significantly reduces the Toffoli-Count and the Toffoli-Depth compared to Jayashree et al.~\cite{jayashree2016ancilla}'s work,  while maintaining the same number of ancilla qubits.\\
    
    In 2022, Sajadimanesh et al.~\cite{sajadimanesh2022practical} extended the applicability of the quantum Shift-and-Add multiplier to real quantum NISQ devices. They achieved this by employing approximate computing, which leverages the resilience of applications to errors and prioritizes efficiency over precision. By developing approximate multipliers that trade accuracy for reduced complexity, the researchers demonstrated that running the proposed approximate multiplier on real quantum computers produces acceptable noise levels for practical applications.\\
    
    \item \textbf{Karatsuba.}\\
    
    The Karatsuba algorithm~\cite{karatsuba1962multiplication} is a fast multiplication algorithm that uses a divide-and-conquer approach to multiple 2 $n$-bit numbers.
    In 2015, Kepley et al.\cite{kepley2015quantum} first used Karatsuba's multiplication algorithm in the quantum domain to develop a software implementation that automatically generates a quantum multiplication circuit. The number of T-gates in this circuit is limited to $7 \cdot n^{\log 3}$ for any given reduction polynomial of degree $n = 2^N$. If an irreducible trinomial of degree $n$ exists, a multiplication circuit with a total gate count of $O(n^{\log 3})$ can be achieved. Compared to benchmarks\cite{Mul_Benchmarks} and schoolbook designs~\cite{parent2017improved, Draper}, this synthesized design uses fewer Toffoli gates and Clifford gates, although it requires more qubits.\\
    
    Two years later, Parent et al.\cite{parent2017improved} introduced an enhanced version of the previous quantum Karatsuba-based integer multiplication. These improvements resulted in a marginal increase in operations by a factor of less than 2 and a slight asymptotic rise in depth for the parallel version. The asymptotic enhancements stem from a thorough analysis of pebble games on complete ternary trees. Specifically, the primary improvement over the previous circuit proposed by Kepley et al.'s\cite{kepley2015quantum} lies in the asymptotic reduction of required qubits, decreasing from $O(n^{1.585})$  to $O(n^{1.427})$, while maintaining a similar Toffoli-Count and Toffoli-Depth.\\
    
    In 2019, Gidney~\cite{gidney2019asymptotically} improved the space complexity of Parent et al.~\cite{parent2017improved}'s quantum Karatsuba multiplication from $O(n^1.427)$ to $O(n)$, while maintaining similar Toffoli complexity. Gidney achieved this by enabling recursive calls to directly contribute their outputs to subsections of the output register, eliminating the need to store and uncompute intermediate results. This technique, similar to classical tail-call optimization, is crucial for optimizing the space complexity of quantum recursive algorithms and can be included in any quantum algorithm design.\\
    
    The following year, Van Hoof~\cite{van2019space} introduced space-efficient variants of the Karatsuba multiplier, requiring 3n qubits. This approach maintains a Toffoli-Count and Toffoli-Depth similar to Kepley et al.~\cite{kepley2015quantum}'s work but incurs a higher number of CNOT gates, theoretically up to $O(n^2)$.\\
    
    In previous studies, the optimization of quantum Karatsuba multiplication has primarily focused on reducing Toffoli-Count or Qubit-Count, neglecting the importance of minimizing Toffoli-Depth. In 2023, Jang et al.~\cite{jang2023quantum} aimed to address this gap by prioritizing the reduction of Toffoli-Depth and full depth in quantum multiplication circuits. Their optimized quantum multiplication approach achieves a Toffoli-Depth of one, significantly lowering the overall depth of the quantum circuit. While their design requires more qubits, it has considerably lower Toffoli-Depth and full depth compared to previous works~\cite{kepley2015quantum, van2019space}.\\

    \item \textbf{Toom-Cook.}\\
    
    The Toom-Cook algorithm is a faster generalization of Karatsuba's method, and several existing quantum implementations are based on this approach.
    \begin{table}[ht!]
    \caption{Asymptotic performance analysis of prominent Toom-Cook quantum multipliers.}
\label{Toom-Cook-Compare}
    \centering
\begin{tabular}{lllll}
\hline
Article              & Structure     & Qubit-Count          & Toffoli-Count        & Toffoli-Depth        \\ \hline
 Putranto et al.~\cite{putranto2023space} & \textcolor{color_toom_cook_2}{Toom-Cook 2}   & $O(n^{1.589})$ & $O(n^{\log_2{3}})$  & $O(n^{1.217})$\\
 Dutta et al.~\cite{Toom-Cook-multi}& \textcolor{color_toom_cook_2_5}{Toom-Cook 2.5} &    $O(n^{1.404})$& $O(n^{\log_6 {16}})$ &$O(n^{1.143})$\\
Larasati et al.~\cite{larasati2021quantum} & \textcolor{color_toom_cook_3}{Toom-Cook 3}   & $O(n^{1.35})$ & $O(n^{2})$ &$O(n^{1.112})$\\
Putranto et al.~\cite{putranto2023space}& \textcolor{color_toom_cook_4}{Toom-Cook 4}  &$O(n^{1.313})$&$O(n^{\log_4 {7}})$&$O(n^{1.09})$\\
Putranto et al.~\cite{putranto2023space}& \textcolor{color_toom_cook_8}{Toom-Cook 8}   &$O(n^{1.245})$ &$O(n^{\log_8 {15}})$ &$O(n^{1.0569})$\\ \hline
\end{tabular}
\end{table}

\begin{figure}[h]
    \centering
    \includegraphics[width=0.72\textwidth]{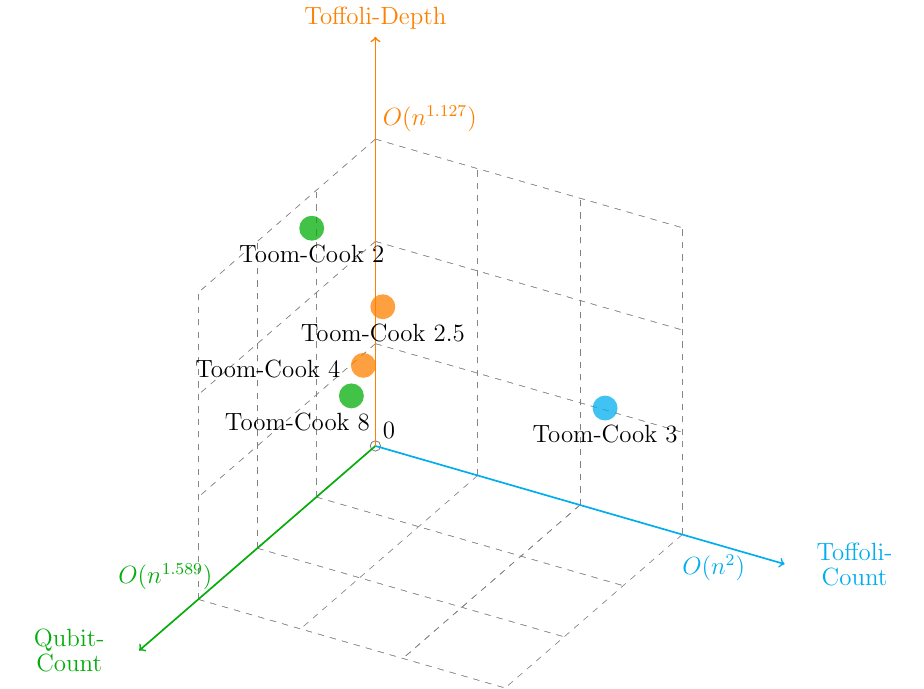}
    \caption{Visualization of the performance of different Toom-Cook quantum multipliers.}
    \label{figToomCook3D}
\end{figure}
    The first quantum Toom-Cook multiplier was proposed by Dutta et al.~\cite{Toom-Cook-multi} in 2018. This initial design is based on Toom-Cook 2.5. By employing reversible pebble games to uncompute the intermediate results, this design was further improved in terms of Toffoli-Depth, Toffoli-Count, and Qubit-Count.\\
    
    In 2021, Larasati et al.~\cite{larasati2021quantum} investigated the quantum circuit for Toom-3 multiplication, which is expected to provide an asymptotically lower depth than the quantum Toom-2.5 design~\cite{Toom-Cook-multi}. Their design achieves lower asymptotic complexity in terms of Toffoli-Depth and Qubit-Count compared to Toom-2.5. However, it involves a larger number of Toffoli gates, primarily due to the implementation of the division operation.\\
    
    In 2023, Putranto et al.~\cite{putranto2023space} proposed a higher-degree multiplier, the Toom-Cook 8-way multiplier, which has the lowest asymptotic performance and implementation cost. To design the Toom-Cook 8-way multiplier, they employed step-by-step computations, including splitting, evaluation, point-wise multiplication, interpolation, and re-composition, along with several strategies to reduce space and time requirements. As shown in Table~\ref{Toom-Cook-Compare}, the Toom-Cook 8-way multiplier demonstrates space- and time-efficient multiplication. More over, Figure~\ref{figToomCook3D} visualizes the performance of different quantum Toom-Cook multipliers.\\
    
    \item \textbf{Wallace Tree.}\\
    
    A multiplier circuit can also be achieved by using the Wallace tree multiplication technique~\cite{wallace1964suggestion}, known as one of the fastest multiplier architectures. This efficient method cascades a sequence of full adders and half adders to sum the partial products.    
    In 2021, Gayathri et al.~\cite{gayathri2021t} proposed the first quantum multiplier based on the Wallace tree multiplication architecture. Compared to the quantum conversion of existing reversible Wallace tree multiplier designs, this work implements a quantum integer multiplication circuit using an efficient quantum full adder, resulting in significant Toffoli-Count and Toffoli-Depth savings.\\
    
    In 2023, Orts et al. \cite{orts2023improving} designed and analyzed a quantum circuit using the Wallace tree technique to compute the product of two integers. Their work involved studying existing circuits in the literature and optimizing the implementation of low-level operations using various techniques. Unlike the quantum Wallace-tree multiplier by Gayathri et al. \cite{gayathri2021t}, which is limited to specific number sizes, this design can handle numbers of any size $N$ and  achieves better Toffoli-Count, Toffoli-Depth, and Qubit-Count.\\
    
    \item \textbf{Sch{\"o}nhage-Strassen.}\\ 
    
    In classical computing, the Sch{\"o}nhage-Strassen algorithm is faster than the Karatsuba and Toom-Cook algorithms, sparking interest in its potential for quantum applications. The first Sch{\"o}nhage-Strassen-based multiplier was designed by Zalka in 1998~\cite{zalka1998fast}, leveraging the $2^{nd}$ level FFT to reduce the computational cost for integer multiplication.\\
    
    In 2023, Nie et al.\cite{nie2023quantum} designed a family of quantum circuits for integer multiplication based on the classical Sch{\"o}nhage-Strassen algorithm. Compared to the quantum Toom-Cook multiplier\cite{Toom-Cook-multi} and the quantum Karatsuba multiplier~\cite{gidney2019asymptotically}, this design offers a reduced gate count and circuit depth but requires more qubits.\\
    
    \item \textbf{Windowed.}\\
    
    Windowed arithmetic, unlike other multiplication methods, aims to reduce the number of operations by using lookup tables to merge operations. 
    In 2019, Gidney~\cite{gidney2019windowed} proposed windowed arithmetic circuits based on lookup tables. Resource estimation conducted using the Azure Quantum Resource Estimator~\cite{hansen2023resource} indicates that the windowed multiplication design offers significant advantages in algorithm runtime and Toffoli requirements.
\end{itemize}

\subsection{QFT-based Circuit Designs}
\begin{figure}[ht!]
    \centering
    \includegraphics[width=0.79\textwidth]{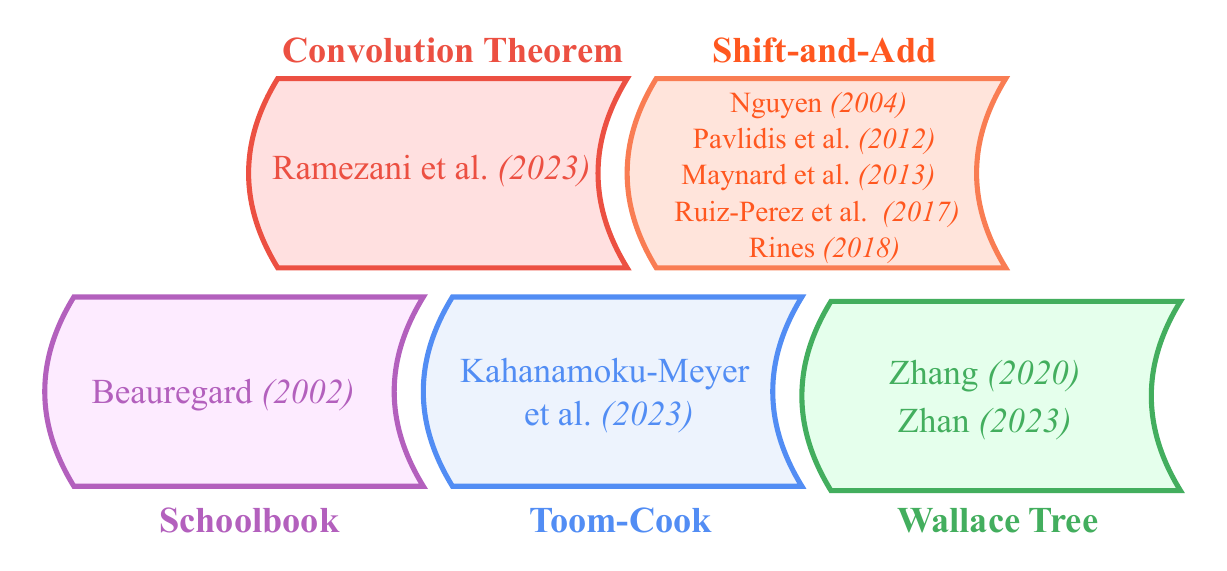}
    \caption{General classification of QFT-based quantum multipliers articles as per structure and sorted by publication year.}
    \label{multiplication_qft}
\end{figure}

Quantum multiplications using QFT-based blocks have been discussed in the literature. Similar to the methods used in Clifford+T-based design, we categorize QFT-based multipliers into the following types, with lines of research listed in Table~\ref{tabQFTMultiplier}. A general classification diagram is provided in Figure~\ref{multiplication_qft} to group each article based on its structure and sorted by publication year for QFT-based quantum multipliers.\\

\vspace*{-12pt}
\begin{table}[!h]
\caption{An overview of prominent QFT-based quantum multipliers.}
\label{tabQFTMultiplier}
\centering
\scalebox{0.8}{
\begin{tabular}{llll}
\hline
Year &Article &Structure & Key Innovations \\
\hline
2002&Beauregard~\cite{beauregard2002circuit}&\textcolor{color_schoolbook}{Schoolbook}&Propose the First Schoolbook Multiplier\\
2004&Nguyen~\cite{nguyen2004tr}&\textcolor{color_shift_and_add}{Shift-and-Add}&Propose the First Shift-and-Add Multiplier\\
2012&Pavlidis et al.~\cite{pavlidis2012fast}&\textcolor{color_shift_and_add}{Shift-and-Add}&Improve Design with Quantum-Classical Adder\\
2013&Maynard et al.~\cite{maynard2014quantum}&\textcolor{color_shift_and_add}{Shift-and-Add}&Propose Multiply-Add Operation \\
2017&Ruiz-Perez et al.~\cite{ruizperez2017quantum}&\textcolor{color_shift_and_add}{Shift-and-Add}&Calculate Controlled Weighted Sum\\
2018&Rines~\cite{rines2018high}&\textcolor{color_shift_and_add}{Shift-and-Add}&Propose Modular Multiplier\\
2020&Zhang~\cite{zhang2020multiplier}&\textcolor{color_problem_aware_designs}{Problem-Aware Designs}&Optimize Depth of Multiplier\\
2023&Zhan~\cite{zhan2023quantum}&\textcolor{color_problem_aware_designs}{Problem-Aware Designs}&Optimize Multiplier with Exponent Encoding\\
2023&Ramezani et al.~\cite{ramezani2023quantum}&\textcolor{color_convolution_theorem}{Convolution Theorem}&Reduce Complexity with Convolution Theorem\\
2023&Kahanamoku-Meyer et al.~\cite{kahanamoku2024fast}&\textcolor{color_toom_cook}{Toom-Cook}&Propose the First Toom-Cook Multiplier\\
\hline
\end{tabular}}
\vspace*{-10pt}
\end{table}
\begin{itemize}
    \item \textbf{Schoolbook.}\\
    
    Similar to Clifford+T-based multipliers, one of the most straightforward implementations of a multiplier is to construct it using sequences of adders. In 2002, Beauregard introduced the first QFT-based multiplier for integer arithmetic with layers of quantum-classical QFT-based adders~\cite{beauregard2002circuit}. While the design is simple, it is inefficient and requires a large amount of resources. Additionally, Beauregard's schoolbook method works only for integer multiplication between a quantum factor and a classical constant.\\
    
    \item \textbf{Shift-and-Add.}\\
    
    Another typical design for a QFT-based multiplier is based on the Shift-and-Add principle. Shift-and-add steps naturally correspond to the binary representation of auxiliary controlled qubits and addition arithmetic. Multiplication of two factors involves setting one factor as the control qubit, which performs additions to the other factor in the Fourier basis. Therefore, multipliers are constructed with controlled QFT-based adders and further decomposed into a QFT block, controlled-controlled phase gates, and an inverse QFT block.\\
    
    The first quantum multiplier based on the Shift-and-Add principle and Draper's quantum-quantum adder was proposed by Nguyen in 2004~\cite{nguyen2004tr}. Pavlidis and Gizopoulos improved it in 2012 by replacing Draper's quantum-quantum adder with Beauregard's quantum-classical adder for multiplying a classical constant to the quantum state~\cite{pavlidis2012fast}. In this work, the structure of the Fourier-based controlled quantum-classical adder was used as a subroutine for constant multiplications.\\
    
    Another novel Fourier-based gate, known as the multiply-add operation, was proposed by Christopher Maynard and Einar Pius in the same year~\cite{maynard2014quantum}. They merged multiplication and addition for classical integers into a single unit consisting of a QFT block, layers of parallel fan-out gates or single-qubit rotation gates, and an inverse QFT block.\\
    
    In 2017, Ruiz-Perez and Garcia-Escart constructed multipliers to calculate controlled weighted sums, extending the applicability for scalable quantum multiplications~\cite{ruizperez2017quantum}. In 2018, Rines improved the QFT-based multiplier to handle modular operations using reduction schemes, including division-based reduction, adaptations of classical Montgomery multiplication, and Barrett reduction~\cite{rines2018high}.\\
    
    \item \textbf{Convolution Theorem.}\\
    
    The convolution theorem states that the convolution of two discrete vectors in the time domain is equivalent to their pairwise multiplication in the frequency domain. This equivalence is quantified by the Fourier transformation of the convolution of two functions and the product of their Fourier transforms.\\
    
    Inspired by this, Ramezani et al. proposed an efficient quantum algorithm for integer multiplication in 2023. By leveraging quantum amplitude amplification, they reduced the time complexity to $\mathcal{O}(\sqrt{n} \log^2 n)$, outperforming the best-known classical algorithm with time complexity $\mathcal{O}(n \log n)$~\cite{ramezani2023quantum}. Since the convolution theorem has no counterpart in Clifford+T-based multipliers, this design is considered a novelty enabled by the power of the Fourier transform. Nevertheless, exploring the application of various theorems related to Fourier transformation that link complex arithmetic to attainable operations remains an open and interesting problem.\\

    \item \textbf{Toom-Cook.}\\
    
    Kahanamoku-Meyer et al. proposed a fast quantum integer multiplication algorithm with zero ancillas based on the classical Toom-Cook algorithm~\cite{kahanamoku2024fast}. They developed an efficient phase rotation algorithm using Draper's QFT-based adder to perform phase product operations and integrated it into the classical Toom-Cook algorithm, achieving sub-quadratic time complexity for multiplications.\\
    
    \item \textbf{Problem-Aware Designs.}\\
    
    Some multiplier designs leverage problem specifications to optimize the circuit extensively. We briefly summarize several intriguingly developed circuits with improved efficiency and reduced applicability, as these designs are problem-aware or based on specific assumptions.\\
    
    For example, in 2023, Junpeng Zhan proposed the Quantum Multiplier based on Exponent Adder (QMbead) as a novel approach for quantum multipliers~\cite{zhan2023quantum}. This approach consists only of an in-place or outplace QFT-based adder. Multiplication results are obtained through measurement outcomes. Although this method is efficient and cost-effective in arithmetic operations, it has a strong restriction on the data encoding procedure. The initial states must be the superposition of all exponents of the two input factors, which is highly expensive in practice~\cite{havlek2019supervised, liu2021rigorous}.\\
    
    Another problem-aware design for QFT-based multipliers involves starting from a conventional quantum multiplier and applying compilation methods to optimize gate depth. An example of this can be found in~\cite{zhang2020multiplier}.
\end{itemize}

\section{Quantum Division\label{sec: div}}
In classical computing, division algorithms can be broadly classified into two categories based on their computational efficiency - fast and slow methods. 
Fast division algorithms are designed to minimize the number of computational steps required for obtaining the quotient and remainder. Two notable examples of such algorithms are the Newton-Raphson and Goldschmidt~\cite{goldschmidt1964applications} which use iterative methods to rapidly converge on a result. 
Conversely, slow division algorithms, such as long division, restoring division, non-restoring division~\cite{shaw1950arithmetic}, and SRT division~\cite{harris1998srt}, are typically inefficient but offer a straightforward approach to understanding and implementation.
Quantum division builds upon the foundational work of classical division, as detailed in the following sections.


\subsection{Clifford+T-based Designs} 
\begin{figure}[ht!]
    \centering
    \includegraphics[width=1.0\textwidth]{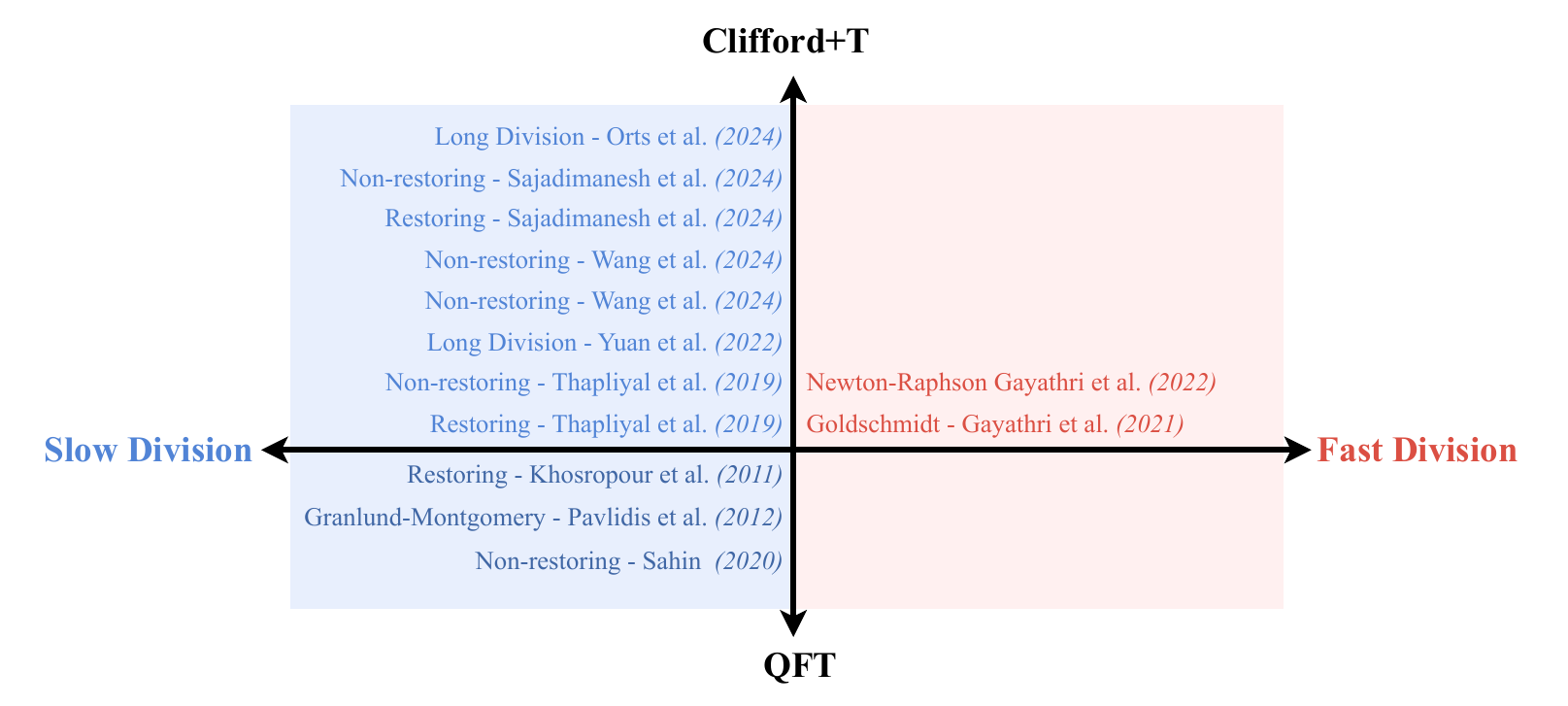}
    \caption{Quadrant analysis chart for quantum dividers based on efficiency and implementation type.}
    \label{division_quadrant}
\end{figure}

Table ~\ref{tabCliffordPlusTDivision} provides a categorization of quantum division using Clifford+T-based quantum dividers, including information on their publication years, articles, structures, and key innovations. To better illustrate each quantum divider's efficiency (fast or slow division) and its implementation type (Clifford+T or QFT), a Quadrant analysis chart is provided in Figure~\ref{division_quadrant}.

\begin{table}[!h]
\caption{An overview of prominent Clifford+T-based quantum dividers.}
\label{tabCliffordPlusTDivision}
\centering
\scalebox{0.9}{
\begin{tabular}{llll}
\hline
Year &Article &Structure & Key Innovations \\
\hline
2019 & Thapliyal et al.~\cite{thapliyal2019quantum}& \textcolor{color_restoring}{Restoring}&Propose the First Restoring Divider\\
2019 & Thapliyal et al.~\cite{thapliyal2019quantum}& \textcolor{color_non_restoring}{Non-Restoring}&Propose the First Non-Restoring Divider\\
2021 & Gayathri et al.~\cite{gayathri2021t}&\textcolor{color_goldschmidt}{Goldschmidt} &Propose the First Goldschmidt Divider\\
2022 & Gayathri et al.~\cite{gayathri2022efficient}&\textcolor{color_newton_raphson}{Newton-Raphson} &Propose the First Newton-Raphson Divider\\
2022 & Yuan et al.~\cite{yuan2022novel}&\textcolor{color_long_division}{Long Division}&Reduce Overall Cost of Designs~\cite{thapliyal2019quantum}\\
2024& Wang et al.~\cite{wang2024boosting} &\textcolor{color_non_restoring}{Non-Restoring}&Apply Comprehensive DSE to Reduce Cost\\
2024& Sajadimanesh et al.~\cite{PhysRevA.109.052601}&\textcolor{color_restoring}{Restoring}&Implement on Quantum Computers\\
2024& Sajadimanesh et al.~\cite{PhysRevA.109.052601}&\textcolor{color_non_restoring}{Non-Restoring}&Implement on Quantum Computers\\
2024& Orts et al.~\cite{orts2024quantum}&\textcolor{color_long_division}{Long Division}&Reduce Toffoli-Count and Toffoli-Depth\\
\hline
\end{tabular}}
\vspace*{-4pt}
\end{table}

\begin{itemize}
    \item \textbf{Slow Division.}\\
    
    In 2019, Thapliyal et al.\cite{thapliyal2019quantum} introduced quantum restoring and non-restoring dividers, inspired by classical slow division algorithms. This marked a significant advancement in the quantum division domain, presenting initial designs based on the Clifford+T set. However, the T-Depth calculation in this work uses a non-standard method, so we disregard the reported cost. The corresponding correction can be found in another work~\cite{wang2024boosting}.\\
    
    Quantum long division remained unexplored until 2022 when Yuan et al.~\cite{yuan2022novel}  introduced a novel fault-tolerant quantum divider based on the long division algorithm using Clifford+T gates, leveraging quantum comparators and subtractors. This design exhibits significant reductions in Toffoli-Count, Toffoli-Depth, and Qubit-Count compared to existing quantum restoring and non-restoring dividers~\cite{thapliyal2019quantum}.\\
    
    Drawing from previous quantum slow division designs, Wang et al.~\cite{wang2024boosting} concentrate on improving the performance of quantum slow dividers by optimizing the design of their sub-blocks. By thoroughly exploring the design space of state-of-the-art quantum addition building blocks, this work achieves notable advancements over previous quantum slow division designs. Moreover, this work underscores the importance of adopting a systematic approach to design space exploration (DSE).\\
    
    In the same year, Sajadimanesh et al.~\cite{PhysRevA.109.052601} used dynamic circuits and approximate computing to implement quantum restoring and non-restoring division circuits on quantum computers. Dynamic circuits were used for mid-circuit measurement, reducing the number of required qubits and enhancing fidelity. Additionally, approximate computing was employed to mitigate noise inherent in quantum hardware. The division circuits were executed on the IBM quantum NISQ computer, demonstrating its capability to overcome quantum noise and yield meaningful results.\\
    
    In 2024, Orts et al.\cite{orts2024quantum} introduced a novel divider circuit based on long division, aimed at reducing the number of T gates while maintaining a comparable Qubit-Count. This approach involves introducing variant minor T-optimized circuits. As a result, the Toffoli-Depth and Toffoli-Count of this design surpass all previous quantum slow division-based dividers\cite{thapliyal2019quantum, yuan2022novel}.\\
    
    \item \textbf{Fast Division.}\\
    
    In 2021, Gayathri et al.~\cite{gayathri2021t} introduced the first quantum fast division circuit based on the Goldschmidt Division algorithm, using the IEEE 754 single-precision format. However, the overall cost of quantum fast division remains significantly higher compared to slow divison designs. 
    One year later, Gayathri et al.~\cite{gayathri2022efficient} proposed a quantum floating-point divider based on the Newton-Raphson method. Table~\ref{compare-fast-division} shows that their division circuit yields substantial savings in T-Count and Qubit-Count compared to state-of-the-art fast division algorithms such as the Quantum Goldschmidt Divider~\cite{gayathri2021t} and the quantum conversion of an efficient reversible floating-point divider based on the Goldschmidt algorithm~\cite{ananthalakshmi2017novel}. This advantage is brought by the improved structure design and more efficient Toffoli decomposition method into T gates.
    \begin{table}[ht!]
    \caption{Performance analysis of prominent fast quantum division designs based on IEEE 754 Single Precision format.}
\label{compare-fast-division}
    \centering
\begin{tabular}{llccc}
\hline
Article              & Structure     & Qubit-Count          & T-Count        & T-Depth        \\ \hline
Gayathri et al.~\cite{gayathri2021t}& \textcolor{color_goldschmidt}{Goldschmidt}&$29074$& $227920$&$59916$\\
AnanthaLakshmi et al.~\cite{ananthalakshmi2017novel}&\textcolor{color_goldschmidt}{Goldschmidt}&$30008$ &$117187$&$17850$\\
Gayathri et al.~\cite{gayathri2022efficient}& \textcolor{color_newton_raphson}{Newton-Raphson}&$23996$&$93376$&$13506$\\
\hline
\end{tabular}
\end{table}
\end{itemize}

\subsection{QFT-based Circuit Designs}
Quantum dividers based on Fourier transformation have characteristics similar to those based on Clifford+T gates. Here we briefly list three exemplars based on slow division algorithms in Table~\ref{tabQFTDivider}. The first QFT-based divider was proposed by Khosropour in 2011 using a restoring division algorithm with QFT-based adders and subtractors~\cite{khosropour2011quantum}. In 2012, Pavlidis used Draper's quantum-quantum adder and Beauregard's quantum-classical structure to develop a quantum divider based on the Granlund-Montgomery classical division by constant algorithm~\cite{pavlidis2012fast}. In 2020, Sahin proposed a non-restoring QFT-based division circuit composed of basic blocks including QFT-based absolute value, QFT-based modular subtractor, two's complement, and modular adder~\cite{sahin2020quantum}.
\vspace*{-12pt}
\begin{table}[!h]
\caption{An overview of prominent QFT-based quantum dividers.}
\label{tabQFTDivider}
\centering
\scalebox{0.9}{
\begin{tabular}{llll}
\hline
Year &Article &Structure & Key Innovations \\
\hline
2011&Khosropour et al.~\cite{khosropour2011quantum}&\textcolor{color_restoring}{Restoring}&Propose QFT-Based Restoring Divider\\
2012&Pavlidis et al.~\cite{pavlidis2012fast}&  \textcolor{color_granlund_montgomery}{Granlund-Montgomery}&Realize Division by a Constant\\
2020&Sahin~\cite{sahin2020quantum}&\textcolor{color_non_restoring}{Non-Restoring} &Propose QFT-Based Non-Restoring Divider\\
\hline
\end{tabular}}
\vspace*{-10pt}
\end{table}


\section{Quantum Modular Exponentiation\label{sec: mod}}
\begin{figure}[ht!]
    \centering
    \includegraphics[width=0.6\textwidth]{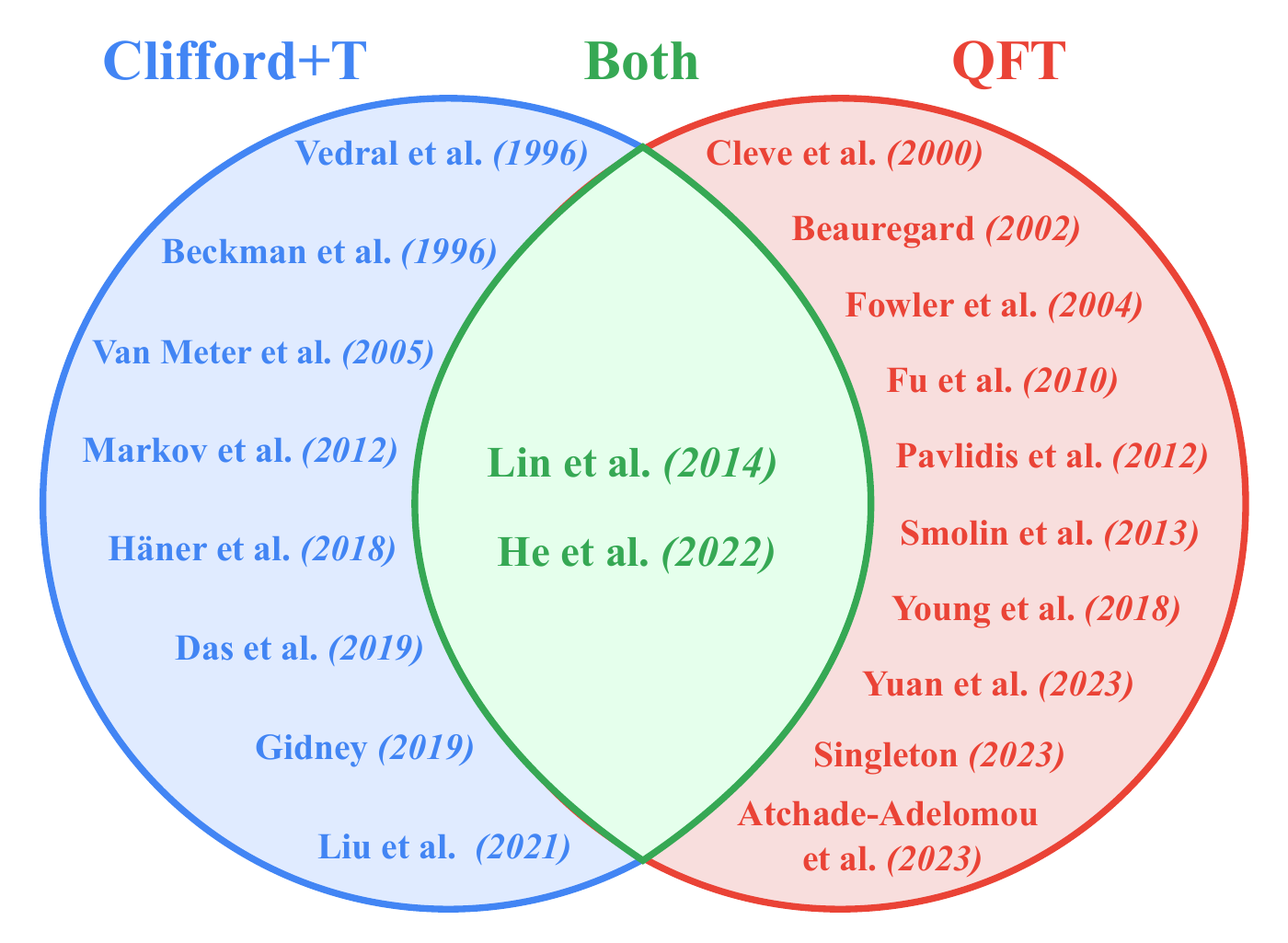}
    \caption{Venn diagram for quantum modular exponentiation designs.}
    \label{modular_exponentiation_venn_diagram}
\end{figure}

Quantum modular arithmetic includes all basic arithmetic operations with bounded variables in a periodic manner. All four basic arithmetic operations - addition, subtraction, multiplication, and division - can be performed in a modular structure. Another infrastructural operation is the modular exponentiation. Although exponentiation itself is a transcendental function that may be complex to implement, quantum algorithms such as Shor's factorization rely heavily on the realization of modular exponentiation. In most architectures, basic quantum arithmetic serves as a fundamental building block for constructing modular exponentiation. Considering the above, this section focuses only on quantum modular exponentiation designs composed of basic modular arithmetic. Unlike the taxonomic approach above, here we put  \textcolor{color_clifford_plus_t}{Clifford+T} designs as well as \textcolor{color_qft}{QFT} designs in~Table~\ref{tableModExp}. Interestingly, we also investigate several works that include both Clifford+T and QFT designs, as listed in~Table~\ref{tableModExp} under \textcolor{color_both}{Both}. For clarity, a Venn diagram for modular exponentiation is provided in Figure~\ref{modular_exponentiation_venn_diagram}.

\begin{table}[!h]
\caption{An overview of prominent quantum modular exponentiation designs.}
\label{tableModExp}
\centering
\scalebox{0.85}{
\begin{tabular}{llll}
\hline
Year &Article &Basis & Key Innovations \\
\hline
1996&Vedral et al.~\cite{vedral1996quantum}&\textcolor{color_clifford_plus_t}{Clifford+T}&Use Arithmetic for Modular Exponentiation\\
1996&Beckman et al.~\cite{beckman1996efficient}&\textcolor{color_clifford_plus_t}{Clifford+T}&Use Arithmetic for Modular Exponentiation\\
2000&Cleve et al.~\cite{cleve2000fast}&\textcolor{color_qft}{QFT}&Implement QFT in Parallel\\
2002&Beauregard~\cite{beauregard2002circuit}&\textcolor{color_qft}{QFT}&Realize Modular Exponentiation with Adder\\
2004&Fowler et al.~\cite{fowler2004implementation}&\textcolor{color_qft}{QFT}&Consider Linear Nearest Neighbour Constraint\\
2005&Van Meter et al.~\cite{van2005fast}&\textcolor{color_clifford_plus_t}{Clifford+T}&Boost the Efficiency with Concurrency\\
2010&Fu et al.~\cite{fu2010speeding}&\textcolor{color_qft}{QFT}&Utilize Semi-Classical QFT\\
2012&Markov et al.~\cite{markov2012constant}&\textcolor{color_clifford_plus_t}{Clifford+T}&Use Predefined Parameters for Specific Cases\\
2012&Pavlidis et al.~\cite{pavlidis2012fast}&\textcolor{color_qft}{QFT}&Implement Using Modular Multipliers and Divider\\
2013&Smolin et al.~\cite{smolin2013oversimplifying}&\textcolor{color_qft}{QFT}&Oversimplify Shor's Algorithm in Experiments\\
2014&Lin et al.~\cite{lin2014qlib}&\textcolor{color_both}{Both}&Propose Benchmarks for Both Bases\\
2018&Young et al.~\cite{young2018simplification} &\textcolor{color_qft}{QFT}&Study Comb Function Using Hadamard Transform\\
2018&H{\"a}ner et al.~\cite{haner2018optimizing}&\textcolor{color_clifford_plus_t}{Clifford+T}&Apply Piecewise Polynomial Approximation\\
2019& Das et al.~\cite{das2019optimizing}&\textcolor{color_clifford_plus_t}{Clifford+T}&Map Logic Designs to Quantum Architectures\\
2019&Gidney~\cite{gidney2019windowed}&\textcolor{color_clifford_plus_t}{Clifford+T}&Introduce Lookup Tables\\
2021&Liu et al.~\cite{liu2021cnot}&\textcolor{color_clifford_plus_t}{Clifford+T}&Minimize the Number of CNOT Gate\\
2022& He et al.~\cite{he2022quantum}&\textcolor{color_both}{Both}&Proposes a BER Protocol\\
2023&Yuan et al.~\cite{yuan2023improved}&\textcolor{color_qft}{QFT}&Design Comparator-Based Modular Exponentiation\\
2023&Singleton~\cite{singleton2023shor}&\textcolor{color_qft}{QFT}&Use Predefined Parameters for Specific Cases.\\
2023&Atchade-Adelomou et al.~\cite{atchadeadelomou2023efficient}&\textcolor{color_qft}{QFT}&Summarize Inplace and Outplace Structures\\
\hline
\end{tabular}}
\vspace*{-4pt}
\end{table}

\begin{itemize}
    \item \textbf{Clifford+T-based Designs.}\\
    
    In 1996, Clifford+T-based quantum modular exponentiation designs emerged. These designs rely on quantum multiplication circuits built from Ripple-Carry adders, which, while simple, are the slowest method.
    As the first work on this topic, Vedral et al.~\cite{VBE} made significant contributions by explicitly constructing quantum networks to perform elementary arithmetic operations, including modular exponentiation. They constructed full modular exponentiation from smaller quantum addition building blocks, with most of the time spent in $20n^2-5n$ Ripple-Carry adders. In this design, the auxiliary memory required grows linearly with the size of the number to be factorized.\\
    
    In the same year, Beckman et al.~\cite{beckman1996efficient} developed a similar full modular exponentiation architecture based on Ripple-Carry adders. Their work differed from Vedral et al.'s by more aggressively leveraging classical computation and presenting several optimizations and trade-offs between space and time.\\
    
    In 2005, Van Meter et al.~\cite{van2005fast} introduced logical concurrency into the existing quantum modular exponentiation framework. By exploiting concurrency, they aimed to reduce the execution time of modular exponentiation, thereby enhancing Shor's algorithm. Their work provided a detailed analysis of the impact of architectural features and concurrent gate execution on quantum modular exponentiation. The main contributions focused on paralleling arithmetic execution through improved adders, concurrent gate execution, and an optimized overall algorithmic structure.\\
    
    Apart from the general modular exponentiation structures capable of handling all possible inputs, one can also create specialized quantum circuits for modular exponentiation with predefined inputs to reduce overall costs. Following this idea, Markov et al.~\cite{markov2012constant} proposed quantum circuits for modular exponentiation in specific input cases in 2012, which provided constant-factor improvements and a significant constant additive term improvement for few-qubit circuits.\\
    
    In 2018, H{\"a}ner et al. implemented modular exponentiation along with other transcendental arithmetic functions such as inverse of \textit{square root}, \textit{sin}, and \textit{arcsin} using piece-wise polynomial approximations~\cite{haner2018optimizing}. They proposed an efficient circuit composed of adder and multiplier to evaluate many polynomials in parallel. The authors referred to adders and multipliers using Clifford+T-based designs, although the blocks are independent of elementary implementations. Therefore, we categorize this as a Clifford+T-based modular exponentiation design.\\
    
    Additionally, mapping practical logic designs to quantum architectures for implementing quantum modular exponentiation is a viable approach. In 2019, Das et al.~\cite{das2019optimizing} used synthesis tools to generate a quantum circuit from a Verilog implementation of exponentiation functions. They converted the logic circuit to a corresponding reversible implementation and mapped it to quantum architectures, demonstrating good efficiency in terms of Qubit-Count, Toffoli-Count, and Toffoli-Depth.\\
    
    In the same year, Gidney~\cite{gidney2019windowed} extended the classical concept of windowing, which uses table lookups to reduce operation counts, to quantum arithmetic. This provided an innovative approach to implementing quantum modular exponentiation. Considering that fault-tolerant quantum computers will be significantly less efficient than classical computers due to noise and expensive error correction, the windowing method is highly efficient in the quantum realm. It trades twenty quantum multiplications for a million classical multiplications, leading to a significant improvement in overall efficiency. Moreover, the proposed design exhibits lower Toffoli-Count than previous works for register sizes ranging from tens of qubits to thousands of qubits.\\
    
    In contrast to previous efforts on minimizing Toffoli-Depth, Toffoli-Count, or Qubit-Count, Liu et al.~\cite{liu2021cnot} aimed to reduce the number of CNOT gates, which significantly impacts the running time on an ion trap quantum computer. Their work achieves basic arithmetic implementations with the lowest known number of CNOT gates. By accumulating intermediate data and employing the windowing technique, they also achieve a low CNOT-count construction for a quantum modular exponentiation design. \\
    
    \item \textbf{QFT-based Designs.}\\
    
    Modular exponentiation circuits using QFT-based blocks have been well discussed in the literature as a key part of Shor's factorization algorithm. Here, we briefly review some of the mainstream works using QFT for modular exponentiation. \\
    
    The most typical model was proposed by Beauregard in 2002~\cite{beauregard2002circuit}. It includes constant number arithmetic operations such as controlled modular adders, controlled modular multipliers, SWAP gates, a QFT block, and an inverse QFT block. In reference~\cite{fowler2004implementation}, Fowler et al. discussed implementing modular exponentiation with the constraint that only linear nearest-neighbor interactions are available. They modified Shor's algorithm by interpolating sequences of SWAP gates and controlled SWAP gates, making it more practical for real hardware settings. Park and Ahn further improved this method for the QFT circuit with linear nearest neighbor architecture by reducing the CNOT count~\cite{park2023reducing}. In 2012, Pavlidis and Gizopoulos proposed the fast quantum modular exponentiation architecture combining quantum-classical modular multipliers and a quantum divider by a constant number~\cite{pavlidis2012fast}. In 2023, Atchade-Adelomou and Gonzalez implemented quantum modular exponentiation with outplace operations using PennyLane software, considering intermediate-scale quantum executions~\cite{atchadeadelomou2023efficient}. In Singleton's work~\cite{singleton2023shor}, the factorization problem is mapped onto another mathematical problem in number theory that finds the period of modular exponential functions. Using the continued fractions algorithm, the complexity of Shor's factorization problem is reduced by directly extracting solutions from approximately measured phase angles, improving the design for the modular exponentiation circuit. \\
    
    An alternative framework for implementing modular arithmetic is based on quantum comparators. A comparator is an oracle that takes two factors, $a$ and $b$, as input and determines whether $a$ is larger than $b$ or vice versa. Modular arithmetic with respect to $a$ and $b$ is then realized by a comparator followed by some basic blocks. For example, in 2023, Yuan et al. designed a novel QFT-based comparator using Beauregard's quantum-classical adder to realize modular addition arithmetic. They then constructed quantum modular multipliers and modular exponentiation circuits~\cite{yuan2023improved}. Although Yuan et, al.'s design requires more QFT blocks compared to the original Beauregard circuit, it reduces the number of auxiliary qubits and improves execution on small-size quantum computers. \\
    
    Additionally, there are several advancements using optimization techniques to reduce the resource cost for QFT and improve modular exponentiation algorithms in practice. For example, in 2000, Cleve and Watrous designed a protocol for fast parallel implementation of QFT by copying Fourier states and estimating phases in parallel, reducing the complexity to logarithmic depth~\cite{cleve2000fast}. In 2018, Young et al. simplified Shor's algorithm for a specific transformation of the comb function and replaced QFT with quantum Hadamard transformation, which requires only controlled rotation gates with angles of $0$ and $\pi$, thus lessening the experimental burden~\cite{young2018simplification}. \\
    
    Another type of optimized modular exponentiation circuits is obtained using semi-classical QFT, proposed by Griffiths and Niu~\cite{griffiths1996semiclassical}. This structure transforms controlled quantum operations followed by measurements into classical controlled operations and post-processing steps, reducing the number of qubits. Inspired by this, Fu et al. utilized semi-classical QFT to speed up the implementation taking into account the non-adjacent form representation~\cite{fu2010speeding}. In 2013, Smolin et al. compiled Shor's factorization algorithm for experimental demonstration~\cite{smolin2013oversimplifying}. Using semi-classical QFT and qubit reuse, they showed both theoretically and experimentally that Shor's algorithm can be realized by small circuits for any composite number $N=pq$. In this complied Shor's algorithm, modular exponentiation is trivial since the exponent only denotes a single qubit and the base is $2$. It is suitable for hardware platforms since it corresponds to the simplest QFT-based adder, made of a Hadamard gate as a single qubit QFT block, a controlled-NOT gate as the controlled phase gate, and another Hadamard gate as a single qubit inverse QFT block. \\
    
    \item \textbf{Adaptable Designs.}\\
    
    In 2014, Lin et al.~\cite{lin2014qlib} proposed Qlib, a quantum module library designed to serve as a suite of benchmarks for quantum logic. The modular exponentiation circuit within Qlib is constructed using a hierarchical decomposition based on a full adder. Interestingly, this work includes both Clifford+T-based and QFT-based quantum modular exponentiation designs, utilizing Cuccaro's Clifford+T-based adder~\cite{Cuccaro} and Draper's QFT-based adder~\cite{draper2000addition}, respectively. In 2022, He et al. designed a quantum modular exponentiation circuit using the binary-exponent-based recombination (BER) protocol~\cite{he2022quantum}. They proposed three implementations of the BER protocol with modular multipliers and modular adders, ranging from Clifford+T-based designs to QFT-based designs.  
\end{itemize} 

\section{Current Applications\label{sec: app}}
Quantum arithmetic plays a crucial role in algorithm implementations. A small improvement in arithmetic design can lead to significant reductions in algorithmic resource cost and computational complexity. Besides modular exponentiation in Shor's factorization algorithm, as discussed in Section~\ref{sec: mod}, quantum arithmetic circuits have wide applications in quantum finance, quantum machine learning, quantum linear systems solvers, and accessing block encoding oracles. In this section, we briefly summarize the potential applications by elaborating an exemplar in each of the four areas. Moreover, we summarize a list of the algorithms and arithmetic kernels in Table~\ref{tableArithmicApplication}.
\begin{table}[!h]
\caption{An overview of prominent applications of quantum algorithms with their quantum arithmetic designs.}
\label{tableArithmicApplication}
\centering
\scalebox{0.8}{
\begin{tabular}{llll}
\hline
Year &Application&Arithmetic Kernel &Algorithmic Design\\
\hline
1996&Cryptography&\textcolor{color_qft}{Modular Exponentiation}&Shor's Factorization Algorithm~\cite{Shor}\\
2009&Linear Systems Solver&\textcolor{color_qft}{Modular Exponentiation}&HHL with Quantum Phase Estimation~\cite{harrow2009quantum}\\
2021&Linear Systems Solver&\textcolor{color_qft}{Modular Exponentiation}&NISQ-HHL with Iterative QPE~\cite{yalovetzky2021hybrid}\\
2021&Financial Modeling&\textcolor{color_both}{Clifford+T and QFT Adder}&Derivatives Pricing Using Various Arithmetic~\cite{chakrabarti2021threshold}\\
2022&Machine Learning&\textcolor{color_clifford_plus_t}{Clifford+T Adder}&Implementation of ReLU with SWAP Gates~\cite{wei2024efficient}\\
2022&Linear Algebra&\textcolor{color_both}{Clifford+T and QFT Adder}&Block-encoding of Classical Data~\cite{clader2022quantum}\\
2022&Linear Algebra&\textcolor{color_both}{Clifford+T and QFT Adder}&Block-encoding of Sparse Matrices~\cite{camps2022explicit}\\
2023&Financical Modeling&\textcolor{color_clifford_plus_t}{Clifford+T Subtractor}&Calculation of Payoff Functions~\cite{lim2023optimized}\\
2023&Linear Systems Solver&\textcolor{color_qft}{QFT-based Adder}&Shift Operators for Hadamard Test~\cite{huang2023hybrid}\\
2023&Linear Algebra&\textcolor{color_both}{Clifford+T and QFT Adder}&Block-encoding of Structured Matrices~\cite{snderhauf2024blockencoding}\\
2024&Machine Learning&\textcolor{color_clifford_plus_t}{Clifford+T Adder}&Constant T-Depth ReLU~\cite{wei2024efficient}\\
2024&Linear Algebra&\textcolor{color_clifford_plus_t}{Clifford+T Adder}&Block-encoding of Banded Matrices~\cite{guseynov2024explicit}\\
\hline
\end{tabular}}
\vspace*{-4pt}
\end{table}

\begin{itemize}
    \item \textbf{Quantum Financial Modeling - Calculating Payoff Functions.}\\
    
    Quantum finance~\cite{ors2019quantum, egger2020quantum, herman2023quantum}, also known as quantum algorithms for computational finance~\cite{rebentrost2018quantumalgorithm, rebentrost2018montecarlo, rebentrost2022martingaleasset}, involves using quantum algorithms to solve classical problems in the financial world with potential advantages. Derivative option pricing is crucial for maximizing returns based on the future price of an underlying asset, such as stocks, currencies, or goods, and a strike price agreed upon in a contract between a buyer and seller. The most typical payoff function is the European call option, which gives the buyer the right, but not the obligation, to exercise the option at the strike price by the expiration date~\cite{black1973pricing, merton1973theory}. The buyer's return is calculated by a nonlinear function that subtracts the strike price from the stock price if the stock price is higher than the strike price, or outputs zero if the stock price is lower. \\
    
    In the quantum world, the dynamics of asset prices are characterized by stochastic processes, which can be predicted using quantum amplitude estimation~\cite{stamatopoulos2020option}, quantum Monte Carlo~\cite{udvarnoki2023quantum}, quantum signal processing~\cite{stamatopoulos2024derivative}, and other methods. However, most algorithms require oracles or subroutines based on fault-tolerant error-corrected computations, resulting in high quantum resource costs, as discussed by Chakrabarti~\cite{chakrabarti2021threshold}. Thus, it remains challenging to develop efficient and cost-effective arithmetic designs for calculating payoff functions in derivative pricing. In 2023, Lim et al. proposed spatially and temporally efficient circuits using quantum adders and subtractors with elementary Cliffford gates and Toffoli gates~\cite{lim2023optimized}. They systematically compared the performance metrics of various quantum adders and used the two's complement method to realize the subtractor. By optimally choosing the best arithmetic kernels, they greatly reduced the resource cost for payoff function calculations and provided efficient options for hardware execution.\\
    
    \item \textbf{Quantum Machine Learning - Implementing Activation Functions.}\\
    
    Quantum machine learning has been a popular topic recently~\cite{biamonte2017quantum, huang2021power, cerezo2022challenges, caro2022generalization, jerbi2023quantum}. Among all the models, the most well-known is the quantum neural network, which takes quantum states as input to the first layer of nodes and evolves to another quantum state after passing through trainable parameterized hidden layers, with information readout via measurements~\cite{kak1995quantum, gupta2001quantum, abbas2021power, nguyen2024theory}. Problem specifications are encoded through the customization of the loss function. Optimization techniques are used to find the optimal parameters, converging to the solutions that correspond to the minimum of the loss function. The transformation from input states to the output states through neural networks is characterized by linear or nonlinear activation functions~\cite{maronese2022quantum}. One of the most common and effective activation functions is the Rectified Linear Unit (ReLU). ReLU is a nonlinear function that outputs the value if it is larger than zero and outputs zero for negative values. \\
    
    In various models of quantum neural networks, the arithmetic design of ReLU usually involves high implementation complexity. In 2022, Sajadimanesh et al. proposed an efficient implementation of the Rectified Linear Unit (ReLU) activation function for an n-qubit state using only one depth of controlled SWAP gates and SWAP gates, making it suitable for NISQ executions~\cite{sajadimanesh2022nisqfriendly}. However, this design requires $n$ auxiliary qubits, which doubles the Qubit-Count. Recently, Zi et al. proposed an efficient circuit design for the implementation of the Rectified Linear Unit (ReLU) with constant T-depth~\cite{wei2024efficient}. They leveraged the quantum fan-out gate and efficient Toffoli gate decomposition to calculate the activation function. They also discussed the implementation of other activation functions using the quantum lookup table, bridging the gap between algorithms and real-world implementations.\\
    
    \item \textbf{Quantum Linear Systems Solver - Realizing Shift Operators.}\\
    
    Another quantum algorithm that can potentially outperform its classical counterpart is the linear systems solver. The most famous quantum algorithm for solving linear systems of equations is the HHL algorithm, proposed by Harrow, Hassidim, and Lloyd in 2009~\cite{harrow2009quantum}. It utilizes quantum phase estimation (QPE) as a subroutine to achieve a quadratic speedup, but it requires a large amount of resources. In 2021, Yalovetzky improved it by using the semi-classical Fourier transformation for iterative phase estimation in the HHL algorithm~\cite{yalovetzky2021hybrid}. In 2019, Huang et al. proposed a near-term version of a linear systems solver with a proof of convergence guarantee~\cite{huang2021nearterm}. This algorithm requires Hadamard tests and convex optimizations to efficiently obtain a classical solution. However, the Hadamard test requires efficient and cost-effective implementation of problem-aware unitaries, limiting the capability of general linear systems solvers with random problem matrices. \\
    
    In solving differential equations, finite difference methods are used to discretize the system into linear systems of equations with banded circulant matrices. This matrix decomposes into a linear combination with powers of parameter shift operators, which have higher complexity in terms of circuit implementation using conventional Clifford+T designs due to the complicated decomposition process~\cite{wilde2009quantum}. Given an $n$ qubit quantum state $\ket{a}=\sum_{i=0}^{N-1}a_i\ket{i}$ and $N=2^n$, a parameter shift operator is a periodic mapping that shifts the computational basis $\ket{i}$ to $\ket{i+1}$ for any $i \in [N]$ and shifts the basis $\ket{N-1}$ back to $\ket{0}$. Inspired by Draper's quantum adder~\cite{draper2000addition}, Koch proposed an efficient scheme for parameter shifting based on QFT and single-qubit rotations~\cite{koch2022gatebased}. This approach avoids the difficulties in Clifford+T design but requires an additional QFT block and its inverse. Based on this design, Huang et al. implemented the parameter shift operators in a Hadamard test using a QFT block and controlled phase gates. They proposed a hybrid quantum-classical algorithm with an improved performance guarantee for solving banded linear systems of equations~\cite{huang2023hybrid}. \\
    
    Although QFT and its inverse are still resource-intensive, we can consider extending the framework of quantum linear systems solvers, previously designed solely for NISQ, to a stage where quantum devices can perform Fourier transform correctly and efficiently~\cite{huang2023hybrid}. This insight can lead to discussions on early fault-tolerant quantum algorithms with randomized Fourier arithmetic, bridging the gap between current hardware platforms and future linear-systems-solver-based applications, including partial differential solvers, matrix inversion, and optimizations~\cite{Katabarwa2023early, yao2023error, kshirsagar2024robust}.\\
    
    \item \textbf{Quantum Linear Algebra - Accessing Block-encoding Oracles.}\\
    
    For most fault-tolerant algorithms, it is essential to equip the data with block-encoding matrices~\cite{snderhauf2024block}. Given an $n$ qubit non-unitary operator $A$ with dimension $N=2^n$, an $m+n$ qubit unitary $U$ is a $(\alpha, m, \epsilon)$-block-encoding of $A$ if $||A-\alpha (\bra{0}^{\otimes m} \otimes I_N) U (I_N \otimes \ket{0}^{\otimes m})||_2\leq \epsilon$~\cite{gilyn2019Quantum}. At first glance, having this special oracle access may seem strange. However due to this specific property, quantum linear and nonlinear algebra can be efficiently implemented using quantum eigenvalue transformation~\cite{low2017oOptimal} or quantum singular value transformation~\cite{gilyn2019Quantum, nonlinear2021guo}. Although the framework seems promising, designing efficient circuits for constructing block-encoding matrices for arbitrarily given matrices remains challenging~\cite{clader2022quantum}. \\
    
    In 2022, Camps et al. proposed an exact circuit construction for well-structured sparse matrices using Clifford+T-based modular adders and QFT-based controlled phase gates~\cite{camps2022explicit}. Later in 2023, Snderhauf proposed an approximate implementation of a block-encoding oracle with in-place additions composed of Clifford+T gates and controlled rotation gates~\cite{snderhauf2024blockencoding}. Furthermore, Guseynov et al. studied the Hamiltonian needed for simulating partial differential equations and proposed an explicit gate construction of block-encoding matrices~\cite{guseynov2024explicit}.
\end{itemize}

\section{Conclusion \label{sec: con}}
In conclusion, the field of quantum computing has significantly evolved in recent decades, yielding remarkable advancements. Quantum arithmetic circuits, fundamental components for constructing quantum algorithms, have garnered considerable attention. While numerous studies on efficient quantum arithmetic circuit designs exist, there has been a pressing need for a systematic review to summarize these achievements.
In this paper, we began with an in-depth discussion on various numerical formats, including integers, fixed-point numbers, and floating-point numbers. We then conducted a comprehensive investigation into quantum arithmetic designs, covering basic operations such as addition, subtraction, multiplication, division, and modular exponentiation. Using various evaluation metrics, we assessed the efficiency of quantum arithmetic circuits based on Clifford+T set and QFT. We also explored potential applications of quantum arithmetic circuits and proposed avenues for future research. Our work provides researchers and engineers with a comprehensive reference to better understand and apply the latest advancements in quantum arithmetic designs, fostering further progress in quantum computing.

In the future, several areas hold promise for advancing quantum arithmetic:
\begin{itemize}
\item \textbf{Windowed Arithmetic:} Most current work focuses on improving arithmetic design circuits, but enhancing lookup table designs also has the potential to advance quantum arithmetic, especially with the rapid development and widespread use of quantum windowed arithmetic~\cite{gidney2019windowed}. Future research could explore the optimization of both quantum arithmetic blocks and lookup tables to achieve better performance.
\item \textbf{Constant Arithmetic:} Compared to variable arithmetic, arithmetic operations involving a variable and a constant, such as reciprocal and squaring, are less explored. While several existing designs, such as those by Markov et al.\cite{markov2012constant} and Singleton\cite{singleton2023shor}, have made contributions, the potential efficiency gains from replacing a variable with a constant remain largely unexplored and represent a valuable area for future research. 
\item \textbf{Arithmetic in Ternary and Other Bases:} In classical arithmetic, multi-valued logic, especially ternary logic, has been extensively explored. Ternary representation in quantum computing could lead to smaller circuit designs with fewer inputs and outputs, reducing the number of qubits needed and enhancing overall efficiency. While most quantum arithmetic designs focus on binary designs, several works have concentrated on ternary reversible designs~\cite{panahi2021novel,asadi2021towards,faghih2023efficient}. There is significant potential in exploring arithmetic in other bases, such as ternary logic. 
\end{itemize}

\vskip6pt
\ack{This research is supported by the National Research Foundation, Singapore under its Quantum Engineering Programme Initiative. Any opinions, findings and conclusions or recommendations expressed in this material are those of the authors and do not reflect the views of National Research Foundation, Singapore.}\vspace{-6pt}
\bibliographystyle{ieeetr}

\vspace{12pt}

\end{document}